\documentclass[sigconf]{acmart}

\AtBeginDocument{%
  \providecommand\BibTeX{{%
    \normalfont B\kern-0.5em{\scshape i\kern-0.25em b}\kern-0.8em\TeX}}}

\copyrightyear{2021}
\acmYear{2021}
\setcopyright{acmlicensed}\acmConference[CHI '21]{CHI Conference on Human Factors in Computing Systems}{May 8--13, 2021}{Yokohama, Japan}
\acmBooktitle{CHI Conference on Human Factors in Computing Systems (CHI '21), May 8--13, 2021, Yokohama, Japan}
\acmPrice{15.00}
\acmDOI{10.1145/3411764.3445526}
\acmISBN{978-1-4503-8096-6/21/05}

\newcommand\revision[1]{{\color{black} #1}}

\usepackage{subcaption}
\usepackage{ulem}

\begin{document}

\title{AutoDS: Towards Human-Centered Automation of Data Science}

\author{Dakuo Wang}
\email{dakuo.wang@ibm.com}
\affiliation{
    \institution{IBM Research}
    \country{USA}
}

\author{Josh Andres}
\affiliation{%
  \institution{IBM Research Australia}
  \country{Australia}
}

\author{Justin Weisz}
\affiliation{%
  \institution{IBM Research}
  \country{USA}}

\author{Erick Oduor}
\affiliation{%
  \institution{IBM Research Africa}
  \country{Kenya}}

\author{Casey Dugan}
\affiliation{%
  \institution{IBM Research}
  \country{USA}}

\renewcommand{\shortauthors}{Wang and et al.}

\begin{abstract}
\revision{

  
  Data science (DS) projects often follow a \textit{lifecycle} that consists of laborious \textit{tasks} for data scientists and domain experts (e.g., data exploration, model training, etc.). Only till recently, machine learning(ML) researchers have developed promising automation techniques to aid data workers in these tasks. This paper introduces \textbf{AutoDS}, an automated machine learning (AutoML) system that aims to leverage the latest ML automation techniques to support data science projects. Data workers only need to upload their dataset, then the system can automatically suggest ML configurations, preprocess data,  select algorithm, and train the model. These suggestions are presented to the user via a web-based graphical user interface and a notebook-based programming user interface. 
  We studied AutoDS with 30 professional data scientists, where one group used AutoDS, and the other did not, to complete a data science project. As expected, AutoDS improves productivity; Yet surprisingly, we find that the models produced by the AutoDS group have \textbf{higher quality} and \textbf{less errors}, but \textbf{lower human confidence scores}. We reflect on the findings by presenting design implications for incorporating automation techniques into human work in the data science lifecycle.
 }
\end{abstract}

\ccsdesc[300]{Human-centered computing~User studies}
\ccsdesc[300]{Human-centered computing~Empirical studies in HCI}
\ccsdesc[300]{Computing methodologies~Artificial intelligence}

\keywords{Data science, automated data science, automated machine learning, AutoML, AutoDS, model building, human-in-the-loop, AI, human-AI collaboration, XAI, collaborative AI}


\maketitle

\section{Introduction and Background}
\begin{figure}
    \centering
    \includegraphics[width=\columnwidth]{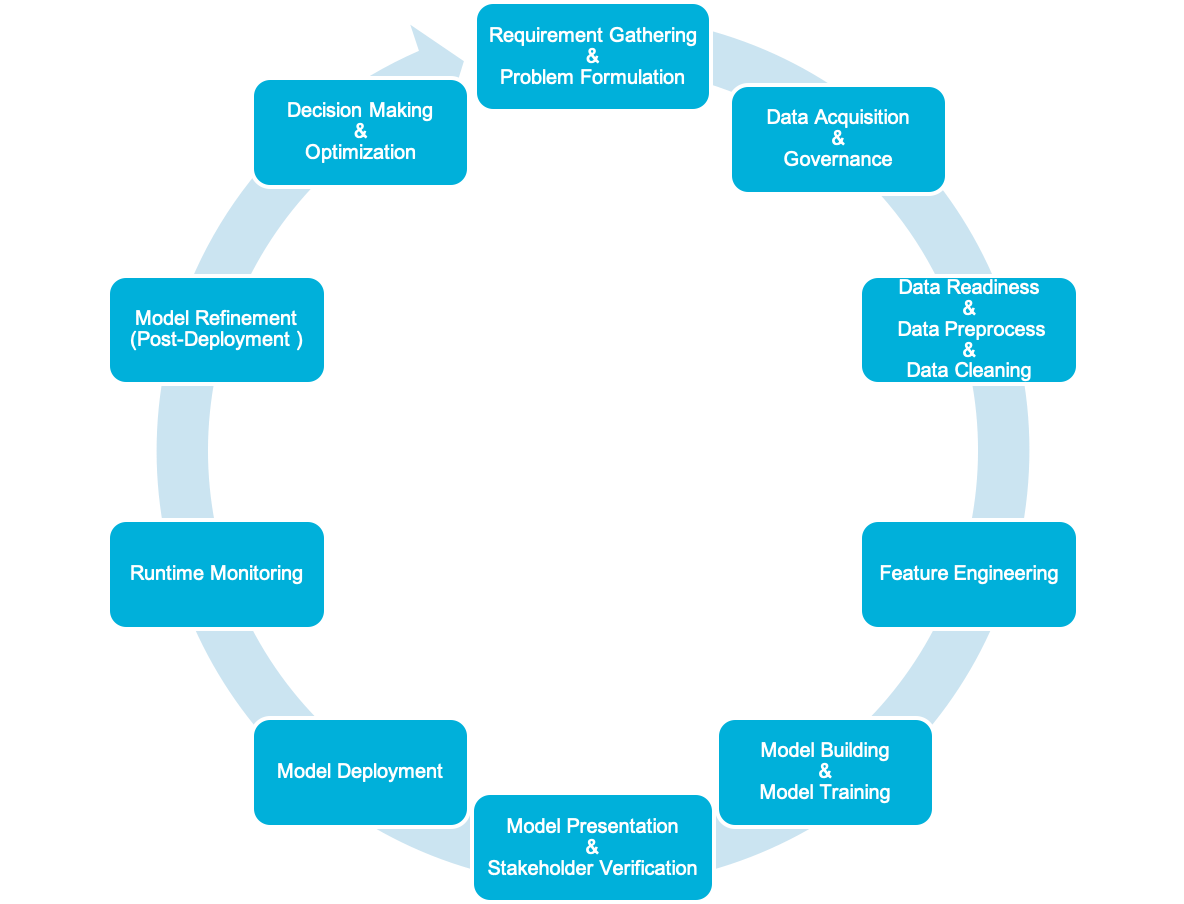}
    \caption{A 10 Stages, 43 Sub-tasks (not shown) DS/ML Lifecycle. This is a synthesized version from reviewing multiple scholarly publications and marketing reports~\cite{wang2019humanai,feurer2015efficient,gartner2020magic,lee2019human,gil2019towards}.}
    ~\label{fig:lifecycle}
\end{figure}

Data Science (DS) refers to the practice of applying statistical and machine learning approaches to analyze data and generate insights for decision making or knowledge discovery~\cite{mao2019,muller2019datascience,donoho201750}. It involves a wide range of tasks from understanding a technical problem to coding and training a machine learning model~\cite{muller2019datascience,wang2019humanai}. 
\revision{Together, these steps constitute a data science project's lifecycle~\cite{automationsurvey}.
As illustrated in Figure~\ref{fig:lifecycle}, data science literature often use a circle diagram to represent the entire data science life cycle.
In one version of the DS lifecycle model, ~\cite{automationsurvey} synthesizes multiple papers and suggests a 10 Stages view of the DS work.
} 
Because the complex natural of a DS lifecycle, a DS project often requires an interdisciplinary DS team (e.g., domain experts and data scientists)~\cite{dscommunicate,zhang2020data, hou2017hacking}. Previous literature suggests that as much as 80\% of a DS team's time~\citep{ruiz201780} is spent on low-level activities, such as manually tweaking data~\citep{heer2012interactive} or trying to select various candidate algorithms~\citep{zoller2019survey} with python and Jupyter notebooks~\cite{web:jupyter}; thus, they do not have enough time for valuable knowledge discovery activities to create better models. To cope with this challenge, researchers have started exploring the use of ML algorithms (i.e. Bayesian optimization) to automate the low-level activities, such as automatically training and selecting the best algorithm from all the candidates~\cite{khurana2016cognito,kanter2015deep,lam2017one,zoller2019survey}; this group of work is called Automated Data Science (or \textbf{AutoDS} for short) \footnote{Some researchers also refer to Automated Artificial Intelligence (AutoAI) or Automated Machine Learning (AutoML). In this paper, we use \textbf{AutoDS} to refer to the collection of all these technologies. }.

Many research communities, universities, and companies have recently made significant investments in AutoDS research, under the belief that the world has ample data and, therefore, an unprecedented demand for data scientists~\cite{he2018amc}. For example, Google released AutoML in 2018~\cite{web:googleautoml}. Startups like H2O~\cite{web:h2o} and Data Robot~\cite{datarobot} both have their branded products. There are also Auto-sklearn~\cite{feurer2015efficient} and TPOT~\cite{web:tpot,olson2016tpot} from the open source communities. 

While ML researchers and practitioners keep investing and advancing the AutoDS technologies,  HCI researchers have begun to investigate how these AI systems may change the future of data scientists' work. For example, a recent qualitative study revealed data scientists' attitudes and perceived interactions with AutoDS systems is beginning to shift to a collaborative mindset, rather than a competitive one, where the relationship between data scientists and AutoDS systems, \revision{cautiously opens the ``inevitable future of automated data science'' ~\cite{wang2019humanai}.}

However, little is known about \textbf{how data scientists would actually interact with an AutoDS system to solve data science problems}. To fill this gap, we designed the AutoDS system and conducted a between-subject user study to learn how data workers use it in practice. \revision{We recruited 30 professional data scientists and assigned them into one of two groups to complete the same task --- using up to 30 minutes to build the best performing model for the given dataset. Half of the participants used AutoDS (experiment group), and the other half built models using Python in a Jupyter notebook, which aims to replicate their daily model building work practice (control group). 
We collected various measurements to quantify 1) the participant's productivity (e.g., how long a participant spent in building the model or improving the model), 2) the final model's performance (e.g., its accuracy), and 3) the participant's confidence score in the final model (i.e., how much confidence a participant has in the final model, if they they are asked to deploy their model). Before the experiment, we also collect measurements on 4) each participant's data science expertise level and 5) their prior experience and general attitude towards AutoDS as control variables.}

\revision{Our results show that, as expected, the participants with AutoDS support can create more models (on average 8 models v.s. 3.13 models) and much faster (on average 5 minutes v.s. 15 minutes), than the participants with python and notebook. More interestingly, the final models from the AutoDS group have higher quality (.919 ROC AUC v.s. .899) and less human errors (0 out of 15 v.s 7 out of 15), than the models from the control group with python and notebooks. The most intriguing finding is that despite participants acknowledged the AutoDS models were at better quality, they had lower confidence scores in these models than in the manually-crafted models, if they were to deploy the model (2.4 v.s. 3.3 out of a 5-point Likert scale). This result indicates that ``better'' models are not equal to ``confident'' models, and the ``trustworthiness'' of AutoDS is critical for user adoption in the future. We discuss the potential explanations for this seemingly conflicted result, and design implications stemming from it.}

In summary, our work makes the following contributions to the CHI community:
\begin{itemize}
    \revision{\item We present an automated data science prototype system with various novel feature designs (e.g., end-to-end, human-in-the-loop, and automatically exporting models to notebooks);
    \item     
    We offer a systematic investigation of user interaction and perceptions of using an AutoDS system in solving a data science task, which yields many expected (e.g., higher productivity) and novel findings (e.g., performance is not equal to confidence, and shift of focus); and
    \item Based on these novel findings, we present design implications for AutoDS systems to better fit into data science workers' workflow.}
\end{itemize}

\section{Related Work}

\subsection{Human-Centered Machine Learning}
Many researchers have studied how data scientists work with data and models. For example, it was suggested that 80 percent of the time spent on a data science project is spent in data preparation  ~\cite{zoller2019survey,muller2019datascience,guo2011proactive,zhang2020data}. As a result, data scientists often do not have enough time to complete a comprehensive data analysis ~\cite{sutton2018data}. 

A popular research topic in this domain is \textbf{interactive machine learning}, which aims to design better user experiences for human users to interact with machine learning tool ~\cite{amershi2014power}. These human users are often labelers of a data sample or domain experts who have better domain knowledge but not so much machine learning expertise. 

Based on these the findings from these empirical studies, many tools have been built to support data science workers' work~\cite{ kandel2011wrangler, rattenbury2017principles,mao2019}. For example, Jupyter Notebook~\cite{web:jupyter} and its variations such as Google Colab~\cite{web:colab} and Jupyter-Lab~\cite{web:jupyterlab} are widely adopted by the data science community. These systems provide an easy code-and-test environment with a graphical user interface so that data scientists can quickly iterate their model crafting and testing process~\cite{kery2018story,muller2019datascience}.  Another group of tools includes the Data Voyager ~\cite{heer2007voyagers} and TensorBoard~\cite{dang2018predict} systems that provide visual analytic support to data scientists to explore their data, but they often stop in the data preparation stage and thus do not provide automated support for model building tasks in the lifecycle (as shown in Figure~\ref{fig:lifecycle}).


\subsection{Human-in-the-Loop AutoDS}
AutoDS refers to a group of technologies that can automate the manual processes of data pre-processing, feature engineering, model selection, etc. ~\cite{zoller2019survey}. Several technologies have been developed with different specializations. For example, Google has developed a set of AutoDS products under the umbrella of Cloud AutoML, such that even non-technical users can build models for visual, text, and tabular data ~\cite{web:googleautoml}. H2O is java-based software for data modelling that provides a python module, which data scientists can import into their code file in order to use the automation capability~\cite{web:h2o}.

Despite the extensive work in building AutoDS systems and algorithms, only a few recent efforts have focused on the interaction between humans and AutoDS. Gil and collaborators propose a guideline for designing AutoDS systems~\cite{gil2019towards}. However, they envision new design features for AutoDS based on their understanding of how data scientists manually build models, and their understandings arise from surveying the previous literature and the authors' personal experience. In our study, we aim to fill in the empirical understanding gap by conduct an experiment to systematically examine how data scientists actually use an AutoDS system.

Another recent work studies data scientists' perceptions and attitudes towards AutoDS systems through an interview with 30 data scientists~\cite{wang2019humanai}. The interviewees, who have never interacted with AutoDS before, believe that AutoDS could potentially change their work practice. They also believe automation in data science work is the future, but they certainly hope that such automation could support their jobs instead of sabotaging them. We follow this line of research. We aim to provide an account for the actual user behaviors when a data scientist uses an AutoDS system to build a model.

There is one more research strand worth mentioning is the information visualization designs for AutoDS systems~\cite{wang2019atmseer,weidele2020autoaiviz}. For example, ATMSeer enables users to browse AutoDS processes at the algorithm, hyperpartition, and hyperparameter levels ~\cite{zoller2019survey}. Their results indicate that users with a higher level of expertise in machine learning are more willing to interact with ATMSeer ~\cite{wang2019atmseer}. We reference these existing visualization work's findings while designing and implementing our prototype system's user interface. But these papers only focus on the visualization aspect and reveal information on existing AutoDS pipelines, and they do not measure the data scientists' behaviors. Thus, the feedback from these studies are limited to the AutoDS visualization user interface design, but not much for the AutoDS's functionality improvement. 

\subsection{Human-Centered AI Design and ``Cooperative'' AI}
The recent "AI Summer"~\cite{grudin2009ai} is remarked with machine learning system demonstrations such as IBM DeepBlue~\cite{campbell2002deep} and Watson Jeopardy~\cite{markoff2011computer}, and Google's AlphaGo~\cite{wang2016does}. In these user scenarios, AI has largely been portrayed as a competitor to humans. The general public and news media began to worry about when the ``singularity'' will arrive, with AI replacing humans~\cite{armstrong2015we}.

More recently, a few researchers have started to argue that instead of worrying about the singularity, why do not we work together to design AI systems that can collaborate with humans?~\cite{li2018goodai, simon_2018} Following this trend,  HCI projects have reported various case studies of designing AI systems to work together with humans instead of replacing them. For example, Cranshaw and collaborators developed a calendar scheduling assistant system that combines the complementary capabilities of humans and AI systems for tasks such as scheduling meetings ~\cite{cranshaw2017calendar}. They coined this architecture a ``human-in-the-loop AI system''. More towards the hardware side of the AI spectrum, a group of researchers from Cornell University have designed a robot that can work together with humans as a team and complete the tasks such as distributing resources to human collaborators ~\cite{claure2019reinforcement}. IBM researchers also experimented the use case of putting an embodied conversational agent into a recruiting team of two human participants, with the agent and humans working together to complete a CV review task ~\cite{shamekhi2018face}. 

There was a seminal debate 20 years ago on ``Agency v.s. Direct Manipulation User Interfaces Design'' ~\cite{shneiderman1997direct}. With more and more deep neural network AI technologies, we, as human users, find it harder and harder to understand what happens inside the AI ``black box''. In parallel, we designate more and more agency and proactiveness to many of today's AI system user interfaces, such as the conversational systems and autonomous cars. Some researchers have reported that users already look at these AI systems differently than the traditional computer systems but more like human partners~\cite{xu161same}, where anthropomorphism effect plays a critical role in user perceptions (e.g.~\cite{nass1994computers, tan2018projecting}. Researchers and designers are actively asking: what are the updated frameworks and theories that we can leverage to help us design better AI systems to work with human? 

Various human-centered design guidelines for AI have been published in the past year from big companies such as Google, Microsoft, and IBM \cite{amershi2019guidelines, schwab_2018, sukis_2019,horvitz1999principles}. But these guidelines often focus only on usability of AI system's interface design, but fall short in discussing the integration of AI systems into human workflows as a collaborative partner. We argue that with an AI system being perceived more like a collaborator teaming up with humans, we may be able to reference classical theories from human-human collaborations to guide our design of an \textbf{cooperative AI system}.  
For example, we may learn from the ``collaboration awareness'' concept ~\cite{dourish1992awareness} to guide our design of AI system's ``transparency''. The ``awareness'' is a bi-directional concept, thus maybe ``transparent AI'' should not only have human better understand AI's runtime state and logic, but also have AI keep track of human's states and intention. 
Thus, in this study, we are also interested in exploring the human-AutoDS interaction through a collaborative work perspective.

\section{AutoDS System Design}

\begin{figure}[h]
\begin{subfigure}{\columnwidth}
\includegraphics[width=\linewidth, height=5cm]{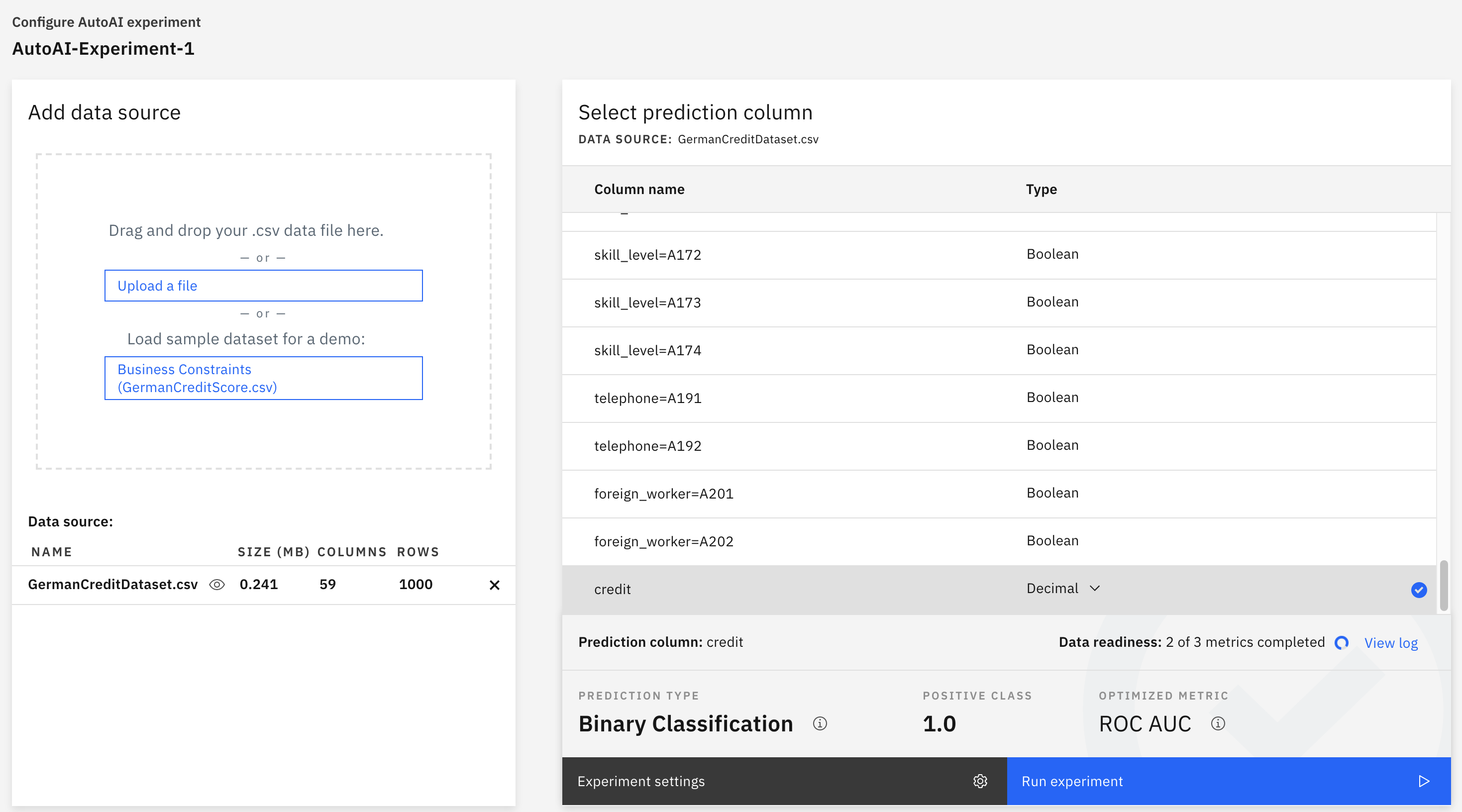} 
\caption{Configuration UI Screen }
\label{fig:subim1}
\end{subfigure}
\begin{subfigure}{\columnwidth}
\includegraphics[width=\linewidth, height=5cm]{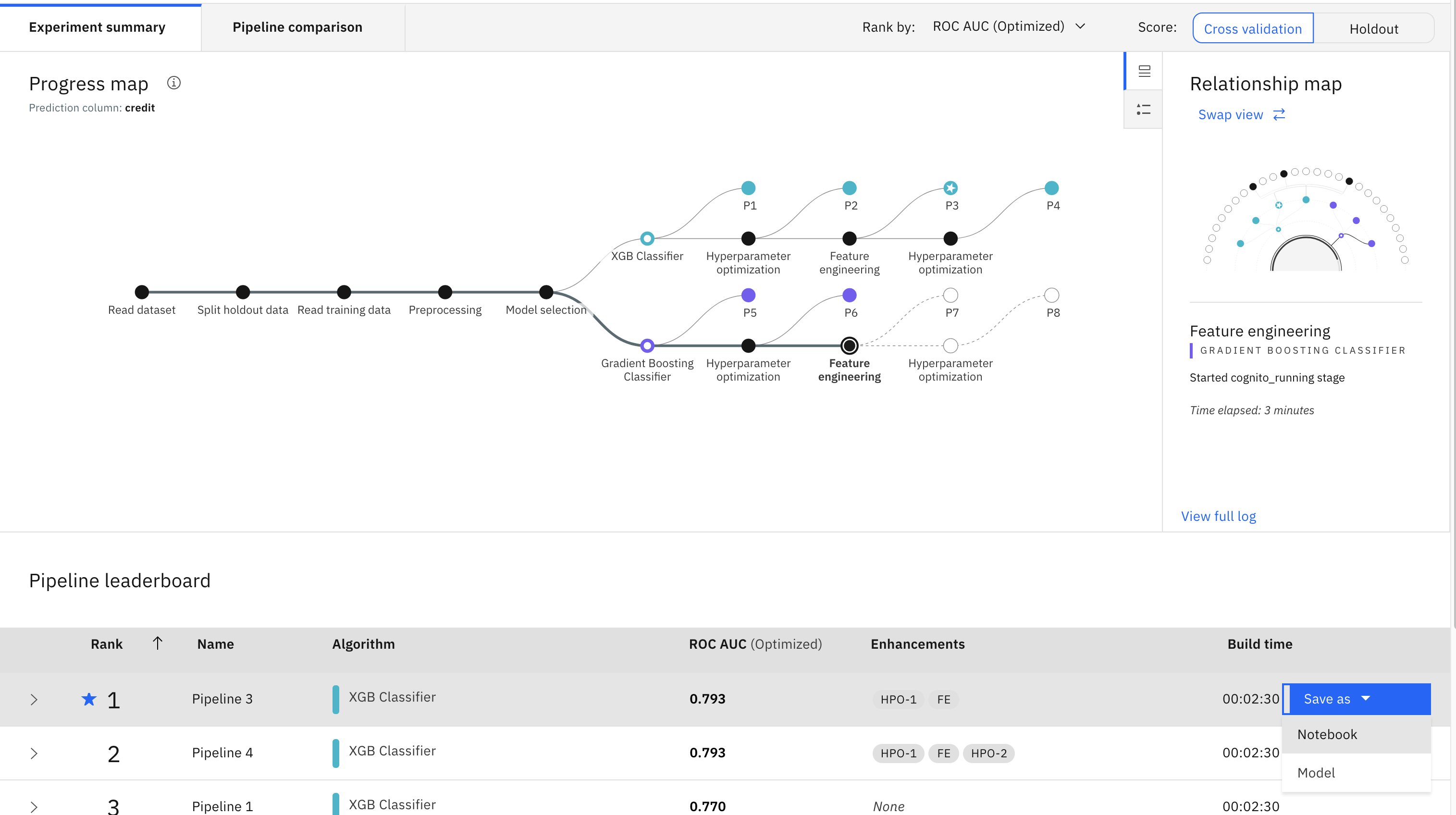}
\caption{Result UI Screen}
\label{fig:subim2}
\end{subfigure}

\caption{The two steps of AutoDS's graphical user interface. }
\label{fig:demo}
\end{figure}

Based on machine learning technique's advances and inspired by design insights from related literature, we implement an AutoDS prototype system, shown in Figure~\ref{fig:demo}, to support data science workers on data science tasks.

From the user perceptive, a user is only required to upload their dataset to trigger the AutoDS execution, and then they can wait for the final model result. They interact with the system primarily from two graphical user interface screens: a \textit{configuration screen (Figure~\ref{fig:subim1})} and a \textit{result screen(Figure~\ref{fig:subim2})}. At the configuration screen, after the user uploads a dataset, AutoDS suggests a specific the ML task configuration for users to approve or adjust (e.g., classification or regression); then at the result screen,  the users can monitor AutoDS's execution progress visualization in real time and eventually see the final results in a model leaderboard view. 

\revision{When looking closer at the result screen in Figure~\ref{fig:subim2}, we design a tree-based progress visualization at the top. Each leave dot in the tree-based visualization represents one of the candidate model pipeline; and the path represents its composition flow. The screenshot illustrates an AutoDS execution is in progress. Two algorithms are being tested (XGB Classifier (blue) and Gradient Boosting Classifier (purple) ). For the XGB Classifier, four models (P1 to P4) are generated; and for Gradient Boosting Classifier, only two models (P5 and P6) are generated, and two more (P7 and P8) are in training. Some models use the same algorithm but their composition steps are different. For example, P2 has an extra step of hyperparameter optimization, in comparison to P1, and P3 has an additional feature engineering step, despite these models are all using the same XGB algorithm.

At the bottom of the result screen in Figure~\ref{fig:subim2}, we included a model result leaderboard. Each row in the leaderboard represents a candidate model pipeline, which corresponds to a model node in the tree visualization at the top. It displays following model information \textit{Rank, Pipeline ID, Algorithm, Accuracy Score, Enhancement Steps such as Transformation and HyperparameterOptimization, Training Runtime}, and \textit{ROC AUC} model performance score on training or holdout data splits.

We design three user functions for each model result, and they all can be triggered through interacting with the model leaderboard. 1) a user can further examine the details of a model (such as the features included or excluded in the model, the prediction plot, etc) through clicking on a row in the leaderboard (not shown in Figure~\ref{fig:subim2}); 2) if a user is satisfied with a model, they can click on the ``Save as'' button to save the ``Model'' (shown in Figure~\ref{fig:subim2}), which will be automatically deployed as a Cloud API endpoint; and 3) if a user is interested in checking the details of a model via python notebook, they can click on ``Save as Notebook'' button (shown in Figure~\ref{fig:subim2}) and download the AutoDS-generated notebook for further improvement. Many of these design consideration (e.g., automatic generation of human-readable python notebooks) are also novel and practical contributions to the AutoDS system designs.

From the system and algorithm perspective, AutoDS reads in the user uploaded dataset, suggests a problem configuration (e.g., classification or regression) based on the data structure.
Then, AutoDS jointly optimizes\footnote{For more technical details about the AutoDS joint optimization algorithm, readers can refer to~\cite{khurana2016cognito,liu2020admm} to replicate the backend.} the sequence of the model pipeline, which includes selecting the appropriate methods for data preprocessing, feature engineering, algorithm selection, and hyperparameter optimization. 
AutoDS can choose a sequence of transformation steps before training an estimator, or it can simply decide not to use any transformation to generate new features. 
Once the sequence is decided, AutoDS can search and decide on which particular transformation to be used inside each of the transformation step, and which estimator to be used inside the modeling step. 
Finaly, AutoDS will fine-tune the hyperparameters of those chosen estimators and transformations simultaneously.

By design, our AutoDS system can provide automation support for the end-to-end data science lifecycle as shown in Fig.~\ref{fig:lifecycle}: from Requirement Gathering and Problem Formulation (i.e., suggesting configurations), to Model Building and Training (i.e., selecting algorithms), and to Decision Making and Optimization (i.e., deploying models).
}

\section{Method}

\subsection{Experiment Design}

\begin{figure}[t]
    \includegraphics[width=\columnwidth]{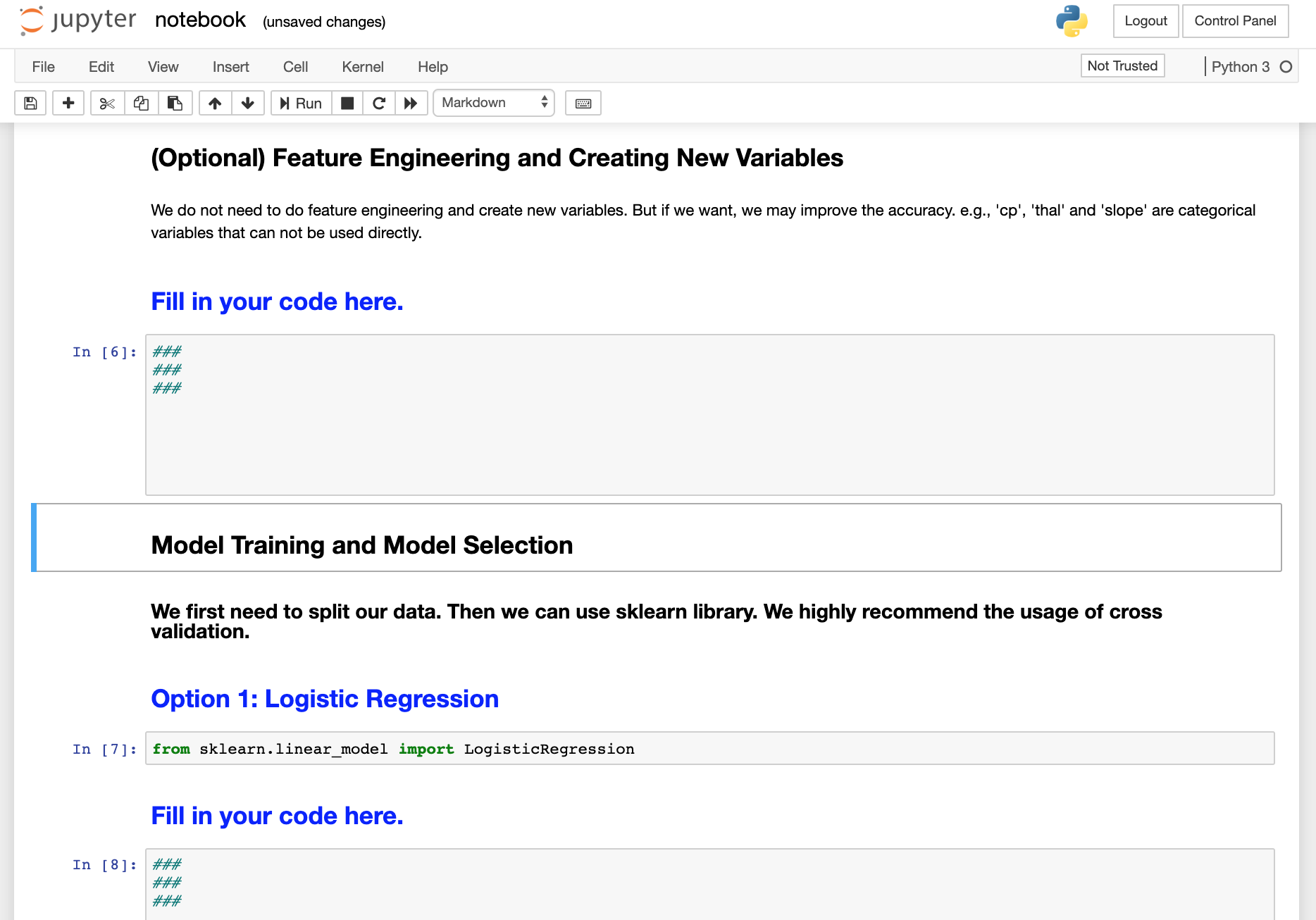}
    \caption{Screenshot from the Jupyter Notebook used in participants group with python and notebook (Control Condition). The Notebook contained skeleton code to load relevant data science libraries (numpy, pandas, skikit-learn), as well as the data set. Instructional sections in the Notebook were provided to remind participants to perform the following tasks: data preprocessing (optional), feature engineering (optional), model training (with 5 different algorithms recommended), and model comparison (optional).}
    \label{fig:notebook}
\end{figure}

Following a lab experiment study guideline~\cite{gergle2014experimental}, we conducted a between-subject user evaluation to understand how people interact with AutoDS in a data science project. 
Participants were asked to try their best at building a machine learning model for the same UCI Heart Disease dataset~\cite{blake1998uci}, either by writing Python code in a Jupyter Notebook (\textbf{Notebook Condition} in Figure~\ref{fig:notebook}) or by using the AutoDS prototype system (\textbf{AutoDS Condition} in Figure~\ref{fig:demo}). 
\revision{The experiment design follows a time-constrained clinical trial setting, each participants was given 30 minutes to build their best model.
The reason why we decided on 30 minutes was because that from our three pilot study sessions, all the pilot participants (data scientists) finished the task within 12 minutes, and they needed to write no more than 10 lines of code to get the expected result: split train and test data subsets, define a model variable, and run a cross-validation function to train the model and report the accuracy score.
We considered a participant having completed the task if s/he got at least one model.

Their final model's performance is evaluated based on the ROC AUC score\footnote{One of the widely-used model performance metrics~\cite{pedregosa2011scikit}. For simplicity, we refer to this metric as ``accuracy'' in the rest of paper.}. 
During the 30 minutes time, participants were allowed to try out one or more models, but they understood that their performance and productivity was not measured by the quantity of the mode, rather the quality of the model.  
They were allowed to submit early if they were satisfied with the model.
All experiment sessions were conducted remotely using a video conferencing system, and participants used their own laptops during the experiment, as both the notebook and AutoDS were hosted on our experiment cloud server.
These requirement was specified in the experiment instruction and all the participants verbally acknowledged this requirement.

Worthy noting that this data science task is a simplified version of their daily data science works, we selected the UCI Heart Disease dataset, which contains only 303 patients' basic medical record information and 13 features (both continuous and categorical values) to predict whether the patients have heart disease (binary target feature with 0 represents no disease). 
It was a widely used benchmark dataset in research papers as well as in machine learning competitions (e.g., Kaggle ~\cite{Kaggle2017}). 
Data cleaning, preprocessing, and feature engineering steps are not critical to build a valid model, but in order to achieve better model performance, participants do need to try out further model improvement steps.
In both conditions, the dataset and required libraries in Notebook were pre-loaded so participants did not need to find the data or set up the network. 
}

We asked the participants to share their screens, and with their consent, we recorded each session for further analysis.

\subsection{AutoDS Condition}
Participants in the \textbf{AutoDS Condition} used our AutoDS prototype, shown in Figure~\ref{fig:demo}. 
As a default problem configuration, AutoDS can generate a list of the four model pipelines with one single algorithm (shown in Fig.~\ref{fig:demo}). 
A model pipeline consists of all of the composition steps in generating the model, from reading data to optimizing hyperparameters.
As aforementioned, not all pipelines had the hyperparameter optimization step or feature engineering step, leading to the variety of generated models and their accuracy scores. 
Participants could modify the AutoDS's configuration and get up to 8 model pipelines.

As shown in the AutoDS system design section, participants have a user function to \textit{inspect} the generated model results with a visualization and a leaderboard showing information such as the confusion matrix, a feature importance chart, and tables with various evaluation metrics (e.g. ROC AUC, precision, F1, etc.) and descriptions of feature transformations. 
In addition, through the automated notebook generation feature in AutoDS systems, participants were also able to download, execute, or edit the generated code and to further \textit{improve} the AutoDS model through coding.

\subsection{Notebook Condition}
Participants in the \textbf{Notebook Condition} were provided with an online Jupyter Notebook environment as a replication to their current work environment. 
To simplify the task, 
we provided notebooks with pre-scripted code sections and a skeleton of instructions. 
The notebooks were preloaded with data and necessary libraries, and we provided instructions in the markdown cells in different sections. 
For example, we listed ``(Optional) Feature Engineering section'', ``(Required) Modeling Training and Selection section'', and blank code cells for users to fill in their code. 
In the model training and selection section, we suggested five commonly used algorithms for this particular task: Logistic Classifier, KNN, SVM, Decision Tree, and Random Forest~\cite{pedregosa2011scikit}. 
\revision{Noted that these pre-scripted code skeletons were meant to help participants to easily write code, and we specifically described in the task sheet that they were not enforced to use these pre-scripted codes. From our observation of the experiment, some participants in the Notebook Condition did choose to not use the skeleton but wrote all codes in one cell.}

\subsection{Measures}

Our research goal was to explore how different the participants (i.e. data scientists) may complete the data science tasks in these two conditions. To that goal, we manually coded the 30 video recording sessions (each is at 45 minutes to one hour long) to extract various measurements, such as how many models each participant generated, what final algorithm they selected, and what its accuracy was. 
We also captured time-related measures, such as how long they spent on the model building task, to evaluate how participants allocate their time differently.


In both AutoDS Condition and Notebook Condition, we collected participants background information and their attitudes and perception towards AutoDS before the session. 
We also collected their confidence score of the final submitted model after the session.

In the AutoDS Condition alone, we also collected participants' perceptions towards the AutoDS prototype they just used through a  XAI questionnaire ~\cite{hoffman2018metrics}, as its four dimensions of predictable, reliable, efficient, and believable of an AI system is applicable in our context.
\revision{In addition, we also collected participants' ratings of trust in AutoDS technologies for a second time in the AutoDS condition, as their scores may change after just experiencing the system.
Thus, we can compare the two pre-study trust scores across the two conditions, and compare the pre-study and post-study trust scores in the AutoDS Condition.

It is worth noting that we collect ROC AUC score among other scores as an indicator of the model performance from both conditions, as we simply need a standardized metric to reflect each participant's task performance.
We acknowledge that the accuracy score of a model should not be overstated~\cite{kay2015good,roy2019automation}, as in an actual data science project, there are various other considerations (e.g., runtime efficiency) in evaluating a model's performance than the model accuracy score.
}

From our observation, some participants in the Notebook Condition made human errors in the submitted model solutions. For example, they forgot to split the data, and thus reported a very high but incorrect accuracy score because they used the training data split.
Other participants reported a F1-score despite the instruction asked for a ROC AUC score.
\revision{Thus, the scores from these error solutions are not comparable to the scores from other participants' solutions.
The participants that made these mistakes acknowledged they were indeed human errors in the post-study interview, and they said they forgot to follow the instruction sheet.
}
To capture this human error, we created one additional measurement to indicate whether a participant successfully complete the task with no human error.

\revision{
In summary, we have four groups of quantitative measurements:
\begin{itemize}
    \item Participant's background information, collected via a pre-study survey (P1, P2 for both conditions);
    \item Participant's perceptions toward general AutoDS type of technologies, collected via a pre-study survey (A1, A2, A3 for both conditions) and post-study survey (AA2, AA3, AA5 for AutoDS condition only)
    \item Participant's behavioral measurements in the task (B3, BNx, BAx) and the final model performance (B1, B2), collected via coding the video recordings (both conditions)
    \item Whether the participant successfully completed the task without human error, collected by examining the final model's code (F1 for both conditions)
\end{itemize}
}
A summary of all the collected quantitative measurements are listed in Table~\ref{tab:measures}. 
We also conducted a semi-structured interview at the end of each session to gather users' qualitative feedback to enrich our results. 

\begin{table}[ht]
\small
\centering
\begin{tabular}{p{8.5cm}}
\hline
\textbf{Background Measures} \\
P1. Years spent practicing data science \\
P2. Prior experience with scikit-learn (1=low, 5=high) \\
\hline
\textbf{Attitudinal Measures} \\
\hspace{0.1cm} \emph{``A'' questions used in both conditions, ``AA'' used only in AutoDS)} \\
A1. Previous familiarity with general AutoDS (1=low, 5=high) \\
A2. Trust in general AutoDS (first time) \\
AA2. Trust in general AutoDS (second time) \\
A3. Belief in general AutoDS replace human (first time) \\
AA3. Belief in  general AutoDS replace human (second time) \\
A4. Participant's confidence in the selected final model \\
AA5. XAI Scale~\cite{hoffman2018metrics}\\
\hline
\textbf{Behavioral Measures} \\
\hspace{0.1cm} \emph{``B'' questions used in both conditions, ``BA'' used only in AutoDS, ``BN'' used only in Notebook} \\
B1. Accuracy of final model (ROC AUC) \\
B2. Type of final model (e.g. logistic regression, random forest, etc.) \\
B3. Total time spent searching for information on the web \\
BN1. Amount of time spent preparing data \\
BN2. Amount of time spent until the first model was produced \\
BN3. Number of different models tried \\
BN4. Did participant perform exploratory data analysis (EDA)? \\
BN5. Did participant perform feature engineering? \\
BA1. Total time spent configuring AutoDS \\
BA2. Total time spent running AutoDS \\
BA3. Total time spent examining the leaderboard \\
BA4. Total time spent examining pipeline details \\
BA5. Number of times AutoDS was run \\
BA6. Number of pipelines examined (i.e. viewed pipeline details) \\
BA7. Number of pipeline notebooks viewed \\
BA8. Number of pipeline notebooks edited \\
BA9. Which AutoDS pipeline was chosen at the end? \\
BA10. Did participant change the code of the final selected pipeline? \\
BA11. Accuracy of the modified pipeline \\
\hline
\textbf{Successful Completion} \\
F1. (Binary) Did participant successfully complete the task without human errors (e.g., mistakenly reporting accuracy score from the training data split instead of the holdout)?  \\
\hline
\end{tabular}
\caption{Summary of background, behavioral, and attitudinal measures captured in the study.}
\label{tab:measures}
\end{table}

\subsection{Participants}
We recruited 30 professional data scientists in a multinational IT company, and randomly assigned them to one of the two conditions (AutoDS v.s. Notebook). These participants come from different locations in the U.S., South Africa, and Australia. Seven participants out of 30 were female (23\%), which was similar to the 16.8\% ratio reported in the Kaggle 2018 survey of data science~\cite{kaggle2018survey}. 

Our recruitment criteria was that the participants were professional data scientists and that they practiced data science or machine learning works in their day-to-day work. Participants reported practicing data science for an average of 3.5 years (SD = 2.7). Six participants (20\%) rated themselves as beginners, 17 (57\%) as intermediates, and 7 (23\%) as experts in data science. 
\revision{Participants rated themselves at a moderate amount of experience (3.2 SD=.97 out of 5) with python scikit-learn lib, which is used in our study. 
Later, in the modeling section in result section~\ref{model-result}, we test whether these background factors (e.g., years of expertise, and prior experience with scikit-learn) may influence the model performance, and the result is not significant.}

\section{Results}
We first describe the overall results, then we list results in the order of: \textbf{attitudes} toward general AutoDS systems; \textbf{behavioral measures} from Notebook conditions and from AutoDS conditions respectively; and lastly, an extensive \textbf{comparison} of the two conditions in terms of participant behaviors, model outcomes, and participant attitudes.

\subsection{Overall Results}
All 15 participants (100\%) in the AutoDS condition and eight participants (53\%) in the Notebook condition finished the model building task without any human error. We refereed to these participants as the ``successful'' participants. In contrast, seven participants (47\%) from the Notebook condition made one or more mistakes in the process. Common human errors included the participant forgetting to split the dataset into training and testing subsets, or reporting a different accuracy metric other than the required ROC AUC score.

The experiment time was limited up to 30 minutes for both the conditions, out of all 30 participants, the fastest participant to reach a state of completion was P23 (Notebook), who finished the task early in 17.9 minutes with an accuracy score of $0.923$. This was an extraordinarily good performance of human data scientists in the Notebook condition, because, on average, participants spent 15 minutes (SD = 6.8 minutes) generating 1 model (BN2). 
For comparison, AutoDS condition participants generated 4 model pipelines in about 5 minutes. 
If the productivity of building a model were to be measured by the time and the quantity, this result could have been seen as ``AutoDS brings \textbf{a 300\% increase in productivity}''. 
\revision{But this result was not the goal of our study and it was kinds of expected. Also, it was not a fair comparison, thus we would caution readers not to use this number bluntly.
}

As for final submitted models, 18 participants chose the Random Forest algorithm (also the best one according to our own experiment), 9 chose Logistic Regression, 2 used SVM and 1 chose ExtraTree. The AutoDS condition outcome was dominated by Random Forest (13 out of 15) with 2 participants chose Logistic regression. In the Notebook condition, participants' choices were more diverse with only 5 out of 15 chose Random Forest. 

\subsection{Attitudes Toward AutoDS}
Before the task, participants in both conditions were asked their familarity and attitudes toward general AutoDS technologies (A1, A2, A3); at the end of the AutoDS condition, participants were again asked about the attitudes questions (AA2, AA3). 
\revision{In this section, we report the comparison between A1, A2, A3 measurements between the two conditions. 
We reserve the comparison result between pre-study trust and post-study trust inside the AutoDS condition in Sec~\ref{sec:trust-result}.}
In general, participants had some familiarity with AutoDS/AutoML technologies (2.27 SD = 1.08 out of 5), but not much. They may have heard about it before or even used it once, but they had not been using it frequently (A1). 

About the ``Trust in General AutoDS/AutoML'' question (A2), we found participants had conflicting opinions: 13 participants (43\%) agreed with this statement, 13 remained neutral, and 4 disagreed (13\%).

And lastly, participants did not believe AutoDS would replace human data scientists (A3): 15 (50\%) disagreed with this statement and only 3 (10\%) agreed, with the rest remaining neutral.

\subsection{How Data Scientists Worked with AutoDS}
In this subsection, we report how participants built models using AutoDS. 

Participants needed to \textbf{explore the data} to understand the problem; they did so in the AutoDS configuration step. 
\revision{Thus, despite AutoDS automatically and instantly suggested default configurations (such as prediction type), participants still spent some time in this step to further examine the dataset by checking the data distributions etc. (2.3 SD=.96 min) (BA1).}

\textbf{Feature engineering, model training, and model selection} steps were fully automated by AutoDS; we tracked time as participants waited for AutoDS to complete runs (2.1 SD=.59 min) (BA2).\footnote{In our study, the dataset was simple, so it took AutoDS only a few minutes to get the first model generated; with a more complicated dataset (e.g., hundreds of MB data file), this training step could take hours.}

\begin{quote} ``It's fast and it gives you visualizations to compare the models, this saves me a lot of time'' (P8, M, AutoDS) \end{quote}

Because it was fast for data scientists in the AutoDS condition to generate models (in total less than 5 minutes), participants had some extra time to spend on other activities (BA5). For example, they spent more time to \textbf{examine or further refine the model} pipeline results generated by AutoDS. They spent most of their time in examining the leaderboard and visualization as shown in Figure~\ref{fig:AutoDS} (9.7 SD=3.4 min) (BA3). Then, they also went into particular model's detailed information page for fine-grained information (3.9 SD=2.2 min) (BA4), 

Participants examined the details of an average of 3.8 (SD=1.2) pipelines (BA6), and some of them downloaded the generated notebook, examined the code to further understand the model (7 out 15 participants, spent an average of 2.3 (SD=.70) minutes) (BA7). Some participants even further revised those codes (7 out 15), and most of these participants edited only one Notebook (BA8). 
Participants who chose to edit code did so with mixed outcome results. 
Out of 7, two were able to improve accuracy scores to the pipeline produced by AutoDS, and two submitted final models that actually had slightly worse accuracy. 
Three abandoned their changes and submitted the original scores (BA10).

The path of selecting the best model in AutoDS condition seemed straightforward: a participant ran AutoDS and selected the model with the ``highest performance AUC ROC score''. But some participants did not select the top performance model as their final choice. Of the 8 possible models that could be produced with AutoDS, 8 participants chose the highest-scoring one (called ``P7'') with a Random Forest algorithm, and 7 chose a different model with a lower ROC AUC score than P7. It appears that even though P7 with the highest ROC AUC score was available to the participant, in some cases they had already invested time and effort in inspecting and tweaking parameters within an earlier generated pipeline, in the end selecting a pipeline with a lower ROC AUC score than P7 (BA9). The mean ROC AUC score for the final selected model was 0.919 (SD=.01) (BA11).

\subsection{How Data Scientists Work with Notebooks}
Now we shift our discussion from user perception to user behavior. Participants' workflow in the Notebook condition was very similar to prior findings, e.g. ~\cite{muller2019datascience}. They went through data exploration, feature generation, model building, hyperparamenter tuning, and modeling selection steps. As this is a lab study centered around model building, participants were not asked to perform deployment steps of the model.

About the time-related features, as aforementioned, participants spent 15 minutes (SD=6.8) in generating the first model (BN2). Note that this BN2 measure also includes the time participants spent in pre-processing and splitting data (BN1 9.4 minutes, SD=5.5), so the actual model building task does not take that long. A few participants (N = 2) were marked ``un-successfull'' as they did not perform the necessary data splitting step (F1). 

In investigating the dataset, seven participants (47\%) performed some form of exploratory data analysis, either by looking at tables that summarized descriptive statistics of the data, or by producing graphs and charts (BN4). But none of the participants wrote new code to explore the data distribution or generate visualizations to support their findings. From the post-study interview, almost all participants mentioned that if they had had more time, they would have loved to conduct more data exploration and understanding of the domain, and consulting with domain experts. By doing so, they believe they could have built better models, and increased their confidence in the model they generated.

\begin{quote}
``... I need to talk with domain experts or doctors to understand the data and feature ... whether my current model makes sense .. Then, I need to think about how to create new features to improve the model.'' (P29, M, Notebook)
\end{quote}

Only four participants (27\%) performed some form of feature engineering, by transforming, scaling, or combining existing features to create new features (BN5). Some argued that they did not have time, while a few others stated they did not do so for this simple, small dataset.
\begin{quote}
``I know SVM's and Random Forest, in problems like this you don't need to pre-normalize [the data], RF can handle it.'' (P28, M, Notebook)
\end{quote}

About the quantity of the outcome, Notebook participants tried on average 3.13 (SD = 1.69) models (BN3). Three participants focused all of their attention on just one model, and four participants tried all five suggested models. Despite finishing the task minutes early, participants who finished only one model argued they did not have enough time to try more models, or so they believed. All of the participants used only the recommended models in the Notebook skeleton, with them justifying:
\begin{quote}
``I would [still] try KNN, RF, Logistic Regression [without your recommendation], because it's a simple binary classification.'' (P18, M, Notebook)
\end{quote}

In summary, the results from the Notebook condition are not surprising. Data scientists followed the general pattern of data science workflow, as reported previously. This is a good result for us, as it means that we have successfully replicated data scientists' day-to-day data science jobs in the lab environment. And by tracking various behavioral and perceptional measures, we now have a solid baseline condition to compare with data scientists' new ways of working with AutoDS. 

\subsection{Human-Crafted Models v.s. AutoDS-Built Models}
We use this section to delve deeply into the comparison of the two ways of working for data scientists in building models. We start with various comparison analyses~\cite{mcdonald2009handbook} on the behavioral patterns from participants, the outcomes of the task, and finally participant attitudes.

\subsubsection{AutoDS Creates More Models Faster and With Fewer Errors}
AutoDS easily generated 8 models with various algorithm and hyperparameter combinations, whereas participants in Notebook condition generated 3.13 (SD = 1.69) models (BN3). To provide a fair comparison, participants in AutoDS examined in detail 3.8 (SD = 1.2) models out of the 8 models (BA6), which served as a candidate set for participants to select the final model. That count (BA6) is still a bit higher than those in Notebook condition (BN3), though the $t$-test shows the difference is not signification $t(28) = 1.25$, $p = .223$. 

About the accuracy of the final selected model, we first reiterate that 7 participants in the Notebook condition reported either an accuracy score other than the ROC AUC metric, or reported the score from the same training data set; as a result, their scores can not be considered valid model scores. We conducted an analysis of variance (ANOVA) to compare the ROC AUC scores (B1) between participants with AutoDS and only the successful Notebook participants. When reporting ANOVA results, we included effect size as partial $\eta^2$, which is the proportion of variance accounted for by each factor in the model, controlling for all other factors. To interpret effect sizes, Miles \& Shevlin~\cite{miles2001applying} gives guidance that partial $\eta^2 \geq .14$ is a large effect, $\geq .06$ is a medium-sized effect, and $\geq .01$ is a small effect. Participants in AutoDS conndition generated and selected the final model with higher ROC AUC scores (.919 SD=.01) than participants with Notebooks (.899 SD=.02), $F [1,21] = 7.61$, $p = .01$, partial $\eta^2 = .27$. (B1)

As aforementioned, participants spent much more time in creating models in Notebook condition than in AutoDS condition. Even if we compare the time participants spent on the first model in Notebook  (time to first model 15.0 SD=6.8 min) (BN2) v.s. the time they generate 8 models in the  AutoDS condition (config + run time 4.3 SD=1.0 min) (BA1+BA2), that difference is significant $t(14.7) = 6.0$, $p < .001$. One factor that contributes to this time difference is that participants in the Notebook condition spent a significantly greater amount of time searching for SK-learn library documentation on the web (6.4 SD=2.6 min) than participants in the AutoDS condition (0.93 SD=1.4) (B3), $t(21.3) = -7.2$, $p < .001$.

Given that participants had limited time in this study, we found that many Notebook participants stopped when they felt they had created a ``good enough'' model and the amount of time they needed to continue to refine their model (e.g. by doing feature engineering) was more than the remaining amount of time in the study. Participants with AutoDS had an easier time understanding \emph{which} model had the highest accuracy, but spent much of their time trying to understand \emph{why}. They leveraged the generated code to interpret the decisions that AutoDS made in the process (BA7), and to evaluate the legitimacy of the AutoDS system. After modifying a couple hyperparameters of the model and seeing the expected changes in accuracy score (BA8), one participant believed they understood why AutoDS selected that particular set of hyperparameters and decided:
\begin{quote}
``Ok, I get an explanation and break down as to what decisions it made, I'll tweak the cross validation then.'' (P4, F, AutoDS)
\end{quote}

\begin{table*}[ht]
    \small
    \centering
    \begin{tabular}{p{6cm}cccccc}
        \textbf{Factor} & \textbf{M} & \textbf{SD} & \textbf{t} & \textbf{p} & \textbf{partial $\eta^2$} & \textbf{Direction} \\
        \hline
            Years spent practicing data science (P1)          & 3.5      & 2.7     & .955   & n.s. & .47   & + \\
            Prior experience with scikit-learn (P2)$\dagger$  & 3.6      & 1.1     & .798   & n.s. & .43   & + \\
            Prior experience with AutoDS (A1)$\dagger$        & 1.7      & .81     & -0.55  & n.s. & .28   & - \\
            Time on task**                           & 29.0 min & 3.6 min & 6.96   & .006 & .73   & + \\
            Time spent searching for information (B3)*   &  .93 min & 1.4 min & -3.19  & .05  & .77   & - \\
            Time configuring AutoDS (BA1)                &  2.3 min & .96 min & 2.17   & n.s.  & < .01 & + \\
            Time examining the leaderboard (BA3)*        &  9.7 min & 3.4 min & 3.51   & .04  & .27   & + \\
            Time examining pipeline details (BA4)*       &  3.9 min & 2.2 min & 4.34   & .02  & .15   & + \\
            Number of pipelines examined (BA6)**          &  3.8     & 1.2     & 7.89   & .004 & .93   & + \\
            Number of pipeline's code inspected (BA7)**   &  2.3     & .70     & 5.33   & .01  & .87   & +  \\
            Number of pipeline's code edited (BA8)    &   .8     & .94     & -1.26  & n.s. & .65   & - \\
    \end{tabular}
    \caption{Factors that predicted higher ROC AUC scores for AutoDS participants. Model adjusted $R^2 = .85$. (* and **) Indicates effects that are both significant ($p \leq .05, \leq .01$) and sizable (partial $\eta^2 \geq .06$). ($\dagger$) Prior experience with scikit-learn and AutoDS were rated on a 5-point scale (1=low, 5=high).}
    \label{tab:success_factors}
\end{table*}

\subsubsection{Factors That Led to More Accurate Models in AutoDS Condition}
\label{model-result}
To understand what factors led to participants choosing a more accurate model while working with AutoDS, we created a regression model to predict ROC AUC scores in AutoDS condition (B1). 
We included a variety of behavioral measures (B3, BA1, BA3, BA4, BA6, BA7, BA8)\footnote{\revision{We added all the behavioral measures except BA5. Because BA5 is the AutoDS computation time, which is a constant number in seconds, thus the model would treat it as a constant and omit it.}} and three self-reported background measures of data scientists (P1, P2, A1) to control for expertise effects: years practicing data science, prior experience with scikit-learn experience, and prior experience with AutDS. We also control how long the participants took to complete the task. For AutoDS participants, we found a number of significant predictors of ROC AUC score, detailed in Table~\ref{tab:success_factors}.

We found that the longer participants searched online for documentation and task-related information in the AutoDS condition (B3), the lower their accuracy score was in the selected model. We observed a similar trend in the Notebook condition, where the more experienced data scientists seemed to spend less time in searching for documentation. 

The longer participants spent in examining the leaderboard and the pipeline details, the better the model they obtained (BA3, BA4). This suggests that the more time and effort participants dedicated to understanding the AutoDS model results, the better their outcome. There were quite a few models generated by AutoDS, and participants only needed to inspect a number of them in detail (BA6). They may have been interested in viewing the pipeline's code in Notebook as well (BA7). Both behaviors led to higher accuracy in the model they selected.

The behavior of changing pipeline code did not necessarily increase the accuracy score (BA8), partially because participants changed the code to make sure they could see corresponding changes in model score, even if the score went down. That gave participants reassurance of the AutoDS generated models.

\subsubsection{Confidence in Human-Crafted Models}
Other than the accuracy score of the final model, we also collected how confident each participant was in their final selected model (A4) at the end of the study. Overall ratings of confidence landed in the middle of the scale (2.9 SD=.97 out of 5), suggesting that participants felt they could have produced even better models if they had had more time. 

Participants in AutoDS condition were significantly less confident (2.4 SD=.98) in their models than participants in the Notebook condition (3.3 SD=.72), $t(25.7) = -2.9$, $p < .01$. An important consideration here is that more confidence did not equate to better models between the two conditions. This may suggest that while the AutoDS users had more options to compare and study, their counterparts in the Notebook condition were invested in writing code from the start. This investment of time and effort into a model could influence users' perspectives when it came to how confident they were. This is backed by the qualitative interview:
\begin{quote}
``I'm pretty confident that this is the best model I can get with this dataset in 30 mins'' (P18, M, Notebook)
\end{quote}

\subsubsection{Trust of AutoDS Increases After Use}
\label{sec:trust-result}
To form a deeper understanding of people's trust in AutoDS, we decomposed the trust into various fine-grain dimensions. We asked AutoDS condition participants to complete an survey developed by Hoffman et al.~\cite{hoffman2018metrics} at the end of the study (AA5). 
The survey was intended to evaluate trust in Explainable AI systems (XAI), and contains 8-items, each with 5-point Likert scale responses ranging from Strongly Disagree to Strongly Agree (Table~\ref{tab:trust_scale}). 
Factor analysis indicated that removing item Q2 increased reliability of the scale, so we aggregated the final scale by omitting this item.

\begin{table}[ht]
    \small
    \centering
    \begin{tabular}{p{8cm}}
         Q1. I am confident in the AutoDS. I feel that it works well. \\
         Q2. The outputs of the AutoDS are very predictable. \\
         Q3. The AutoDS is very reliable. I can count on it to be correct all the time. \\
         Q4. I feel safe that when I rely on the [AutoDS] I will get the right answers. \\
         Q5. AutoDS is efficient in that it works very quickly. \\ 
         Q6. I am wary of the AutoDS. \\
         Q7. AutoDS can perform the task better than a novice human user. \\
         Q8. I like using the system for decision making. \\
    \end{tabular}
    \caption{XAI survey from ~\cite{hoffman2018metrics}. With the removal of Q2, this scale has acceptable reliability (Cronbach's $\alpha = 0.75$).}
    \label{tab:trust_scale}
\end{table}

Previous literature~\cite{hoffman2018metrics} suggests to use the XAI scales as an overall score. The overall XAI score was quite positive (3.6 SD=.41 out of 5) (AA5). Participants found the AutoDS efficient, predictable, and liked it for decision making. 

\begin{quote} ``It's fast and it gives you visualisations to compare the models, this saves me a lot of time'' (P8, M, AutoDS) \end{quote} 

However, participants wondered about the reliability of AutoDS, trying to understand the rationale behind AutoDS's decisions. \begin{quote}``I'm not sure about the features it created for me'' (P5, M, AutoDS), ``It selected in one case random forest and in another logistic regression, but why, can i choose the classifier?'' (P6, F, AutoDS) \end{quote}

In order to gauge how participants felt about AutoDS after having used it, we measured their attitude toward AutoDS twice, once at the beginning of the study (A2, A3) and once at the end (AA2, AA3). Participants' opinions generally did shift in a positive direction after experiencing AutoDS. Participants had significantly higher levels of trust in AutoDS after using it ( 3.5 SD=.74) than before (2.7 SD=1.1), $t(24.4) = -2.3$, $p = .03$. 
This finding is consistent with the finding that all 15 participants reported that they want to adopt AutoDS in their day-to-day model building work after the experiment. 
As to whether our participants felt that AutoDS would one day replace data scientists, there was a small shift in opinion toward feeling that it would, but it still falls in the negative direction, and this difference was not significant (A3) (pre-study 2.4 SD=.83,  post-study 2.9 SD=1.2, $t(25.3) = -1.4$, $p = n.s.$).

\section{Discussion}

Motivated by literature, we began the investigation on how data scientists use an AutoDS system in DS tasks. Through our analyses, we found that participants working with AutoDS not only can create more models and faster, but more importantly these models are at a higher accuracy and with fewer human-errors, comparing to participants working with a Jupyter Notebook coding environment. With AutoDS's help, data scientists can afford to spend more time in understanding the results and inspecting the codes generated by AutoDS, whereas in the baseline condition participants had to spent lots of their efforts and time in searching for library documentations and writing code from scratch. 
\revision{However, despite participants acknowledge all these benefits from using AutoDS, participants still trust their manually-crafted models more than the AutoDS generated model. 
These results suggest design improvement for AutoDS technology, and enlighten us about the future of human-centered automated data science work. 

\subsection{Design Implications}
Our results suggested that AutoDS prototype's code generation feature, which automatically translate a resulting model into human-readable Python code, is very welcomed by the participants. They could inspect the codes to see the decisions made by AutoDS. They can rationalize some of those decisions, and they could even change the code and see if that changes the result as expected. Furthermore, we observed there are a couple participants who kept working on the code generated by AutoDS to achieve a better model, that use case again supports our claim that data scientists and AutoDS together could deliver better results. Other than the possible use of the entire code, data scientists could also easily migrate or replicate a segment of the generated code, maybe with a transformation function, some hyperparameters, or other decisions made by AutoDS into a different task and context. Thus, we highly recommend AutoDS systems to provide such code-generation feature.

We also learned that participants have trouble in trusting AutoDS system, and they have more confidence in their worse-performing manually-created models than in the higher quality AutoDS-generated models, thus the explainable and trustworthy AutoDS research agenda should be prioritized. The fact that people loved the code-generation feature could also attribute to the needs of a transparent and trusted AutoDS. Participants wanted to understand not only what the decisions are (code-generation satisfied that), but also why AutoDS made such decisions. However, only the generated code for the result model was not enough. We need to design AutoDS systems to provide better explanation for why certain decisions were made. For example, when showing users the top performed algorithm and the optimized set of hyperparameter values, we could also reveal the information about all the other candidates considered, and why AutoDS eliminated those options. Another example is that for the new features engineered by AutoDS, users should have a better view of how those features are generated and chosen in the middle of the feature engineering process, because users may have some domain knowledge to prevent the nonsense transformations (e.g., absolute value of gender). We are glad to see that recently there are some researchers moving toward this direction~\cite{JaimieDrozdal2020}. We should also caution readers that we are not arguing to show a user all the information AutoDS has. Full transparency causes information overload and is no better than no transparency. This links back to a well-established CSCW theory: ``social translucent''~\cite{erickson2000social}, where researchers argue to show the information at the right moment and at the right level of details, and we should learn from our past.

\subsection{Human-Centered Automated Data Science}

Our result showed that some participants worked over a wider range of AutoDS results, some others focused on improving the code of one particular \revision{model (BN6 and BA7). The more successful }participants were the ones who tried a wider range of approaches rather than focusing all of their attention on just one model. 
This finding echoed wisdom in the artistic domain, where the most creative works emerge when favoring quantity over quality~\cite{bayles2001art}. And this was not isolated that the data science work appears linked with a practice in the art design domain, as ~\cite{muller2019datascience} argued that ``Data science workers work is a new type of crafting'', while in the AutoDS context, users are crafting the AutoDS generated results as a material.

This collaborative step that emerged over practice, between data scientist and the AutoDS, appears to us as a \textit{demythification} of the AutoDS replacing the job of the data scientist, and instead, offers a perspective on how the relationship between this parties towards a collaborative partnership is maturing. We argue that the \textbf{Human-AI Collaboration}~\cite{wang2019humanai} paradigm is also emerging in the data science domain, as in many other domains (e.g., education~\cite{xu161same}, healthcare~\cite{aidoctor}, human resource~\cite{shamekhi2018face}). 
In the data science context, human data scientists can work on the understanding of domain knowledge and context tasks, while AutoDS systems work on the computation tasks. Then human data scientists can work on result interpretation tasks, while AutoDS can work on deployment tasks. They hand over information to each other and share responsibility. Together, AutoDS plays a partner role in the future of human-centered automated data science practice.

We believe AutoDS technologies will become available to more and more data scientists, affording them a chance to craft models together with AI. 
We agree with the majority of our participants that AutoDS will not replace data scientists' job. 
Instead, a human-centered AutoDS will play a more important assistant role to the human data scientists in new paradigm of data scietist's work.

In this new paradigm, data scientists' role will shift, as suggested by our findings: they can spend less time in the tedious model training and hyperparameter tuning tasks, but spend more time in understanding the data, communicating with the domain experts, and selecting the model to best fit the domain and context. In addition, they will need to spend a big proportion of their time to rationalize the AutoDS decisions and results, and translate those decisions to present to other human stakeholders. 

In this new paradigm, not only professional data scientists but also end-customers and domain experts can also build ML solutions to answer their own simple questions with their data. Our system provide a GUI and an automatically export human-readable notebook in automatically models building. 
The future in which various human stakeholders can work together with AutoDS is promising, but there are works needed to be done to achieve that vision. This work is just the first step.


}

\subsection{Limitations}
We acknowledge our works limitations as follows: this is a lab experiment study, thus it has the common limitations as any other lab experiments~\cite{gergle2014experimental}. 
\revision{
For example, in the Notebook Condition, we provided participants a code skeleton, which may also bias their behaviors in that condition.
Another example is that the task and data set is tailored as simple and specific tasks to emphasize model building, thus the reported behaviors of participants interacting with AutoDS may be different from when they actually adopt AutoDS in their daily DS projects.
In an actual data science project, data acquisition and curation are more significant problems for data scientists during the ML lifecycle. 
The AutoDS system presented in this paper is just the starting point of a long-term project. We have another paper in submission that discusses human-AI collaboration in automated data preparation and curation process. In that work, we used reinforcement learning instead of AutoML, because it is better suited for the task.
}

Another limitation is that all the participants are professional data scientists coming from a technology company, there may be some selection bias in their background. 
\revision{For example, on average, they rated themselves as moderate level of data science expertise. 
This participant population may also impose limitations on the generalizability of our paper's findings.
We call out these external validity limitations to the reader who plan to apply the findings to other contexts.
}

\section{Conclusion}
Our work presents an AutoDS prototype system, and a between-subject user experiment with with 30 professional data scientists to use the AutoDS system, or use python in notebooks, to complete a data science model building task. Grounded in the results, we present design implications for building AutoDS systems to better incorporate into human workflows. We also discuss future research directions for human-centered automated data science.

\begin{acks}
We thank all the participants who contributed their time participate our experiment. We want to specifically thank Arunima Chaudhary, Abel Valente, and Carolina Spina for implementing the presented AutoDS system. We also want to thank Alex Swain, Dillon Eversman, Voranouth Supadulya, and Daniel Karl for their inspiring UX design sketches, and Gregory Bramble, Theodoros Salonidis, Peter Kirchner, and Alex Gray for their input on the algorithm implementation.
\end{acks}

\bibliographystyle{ACM-Reference-Format}
\bibliography{sample-base}


\begin{thebibliography}{71}


\ifx \showCODEN    \undefined \def \showCODEN     #1{\unskip}     \fi
\ifx \showDOI      \undefined \def \showDOI       #1{#1}\fi
\ifx \showISBNx    \undefined \def \showISBNx     #1{\unskip}     \fi
\ifx \showISBNxiii \undefined \def \showISBNxiii  #1{\unskip}     \fi
\ifx \showISSN     \undefined \def \showISSN      #1{\unskip}     \fi
\ifx \showLCCN     \undefined \def \showLCCN      #1{\unskip}     \fi
\ifx \shownote     \undefined \def \shownote      #1{#1}          \fi
\ifx \showarticletitle \undefined \def \showarticletitle #1{#1}   \fi
\ifx \showURL      \undefined \def \showURL       {\relax}        \fi
\providecommand\bibfield[2]{#2}
\providecommand\bibinfo[2]{#2}
\providecommand\natexlab[1]{#1}
\providecommand\showeprint[2][]{arXiv:#2}

\bibitem[\protect\citeauthoryear{Amershi, Cakmak, Knox, and Kulesza}{Amershi
  et~al\mbox{.}}{2014}]%
        {amershi2014power}
\bibfield{author}{\bibinfo{person}{Saleema Amershi}, \bibinfo{person}{Maya
  Cakmak}, \bibinfo{person}{William~Bradley Knox}, {and} \bibinfo{person}{Todd
  Kulesza}.} \bibinfo{year}{2014}\natexlab{}.
\newblock \showarticletitle{Power to the people: The role of humans in
  interactive machine learning}.
\newblock \bibinfo{journal}{\emph{AI Magazine}} \bibinfo{volume}{35},
  \bibinfo{number}{4} (\bibinfo{year}{2014}), \bibinfo{pages}{105--120}.
\newblock


\bibitem[\protect\citeauthoryear{Amershi, Weld, Vorvoreanu, Fourney, Nushi,
  Collisson, Suh, Iqbal, Bennett, Inkpen, et~al\mbox{.}}{Amershi
  et~al\mbox{.}}{2019}]%
        {amershi2019guidelines}
\bibfield{author}{\bibinfo{person}{Saleema Amershi}, \bibinfo{person}{Dan
  Weld}, \bibinfo{person}{Mihaela Vorvoreanu}, \bibinfo{person}{Adam Fourney},
  \bibinfo{person}{Besmira Nushi}, \bibinfo{person}{Penny Collisson},
  \bibinfo{person}{Jina Suh}, \bibinfo{person}{Shamsi Iqbal},
  \bibinfo{person}{Paul~N Bennett}, \bibinfo{person}{Kori Inkpen},
  {et~al\mbox{.}}} \bibinfo{year}{2019}\natexlab{}.
\newblock \showarticletitle{Guidelines for human-AI interaction}. In
  \bibinfo{booktitle}{\emph{Proceedings of the 2019 CHI Conference on Human
  Factors in Computing Systems}}. ACM, \bibinfo{pages}{3}.
\newblock


\bibitem[\protect\citeauthoryear{Armstrong and Sotala}{Armstrong and
  Sotala}{2015}]%
        {armstrong2015we}
\bibfield{author}{\bibinfo{person}{Stuart Armstrong} {and} \bibinfo{person}{Kaj
  Sotala}.} \bibinfo{year}{2015}\natexlab{}.
\newblock \showarticletitle{How we’re predicting AI--or failing to}.
\newblock In \bibinfo{booktitle}{\emph{Beyond artificial intelligence}}.
  \bibinfo{publisher}{Springer}, \bibinfo{pages}{11--29}.
\newblock


\bibitem[\protect\citeauthoryear{Bayles and Orland}{Bayles and Orland}{2001}]%
        {bayles2001art}
\bibfield{author}{\bibinfo{person}{David Bayles} {and} \bibinfo{person}{Ted
  Orland}.} \bibinfo{year}{2001}\natexlab{}.
\newblock \bibinfo{booktitle}{\emph{Art \& fear: Observations on the perils
  (and rewards) of artmaking}}.
\newblock \bibinfo{publisher}{Image Continuum Press}.
\newblock


\bibitem[\protect\citeauthoryear{Blake and Merz}{Blake and Merz}{1998}]%
        {blake1998uci}
\bibfield{author}{\bibinfo{person}{Catherine~L Blake} {and}
  \bibinfo{person}{Christopher~J Merz}.} \bibinfo{year}{1998}\natexlab{}.
\newblock \bibinfo{title}{UCI repository of machine learning databases, 1998}.
\newblock
\newblock


\bibitem[\protect\citeauthoryear{Campbell, Hoane~Jr, and Hsu}{Campbell
  et~al\mbox{.}}{2002}]%
        {campbell2002deep}
\bibfield{author}{\bibinfo{person}{Murray Campbell}, \bibinfo{person}{A~Joseph
  Hoane~Jr}, {and} \bibinfo{person}{Feng-hsiung Hsu}.}
  \bibinfo{year}{2002}\natexlab{}.
\newblock \showarticletitle{Deep blue}.
\newblock \bibinfo{journal}{\emph{Artificial intelligence}}
  \bibinfo{volume}{134}, \bibinfo{number}{1-2} (\bibinfo{year}{2002}),
  \bibinfo{pages}{57--83}.
\newblock


\bibitem[\protect\citeauthoryear{Claure, Chen, Modi, Jung, and
  Nikolaidis}{Claure et~al\mbox{.}}{2019}]%
        {claure2019reinforcement}
\bibfield{author}{\bibinfo{person}{Houston Claure}, \bibinfo{person}{Yifang
  Chen}, \bibinfo{person}{Jignesh Modi}, \bibinfo{person}{Malte Jung}, {and}
  \bibinfo{person}{Stefanos Nikolaidis}.} \bibinfo{year}{2019}\natexlab{}.
\newblock \showarticletitle{Reinforcement Learning with Fairness Constraints
  for Resource Distribution in Human-Robot Teams}.
\newblock \bibinfo{journal}{\emph{arXiv preprint arXiv:1907.00313}}
  (\bibinfo{year}{2019}).
\newblock


\bibitem[\protect\citeauthoryear{Cranshaw, Elwany, Newman, Kocielnik, Yu, Soni,
  Teevan, and Monroy-Hern{\'a}ndez}{Cranshaw et~al\mbox{.}}{2017}]%
        {cranshaw2017calendar}
\bibfield{author}{\bibinfo{person}{Justin Cranshaw}, \bibinfo{person}{Emad
  Elwany}, \bibinfo{person}{Todd Newman}, \bibinfo{person}{Rafal Kocielnik},
  \bibinfo{person}{Bowen Yu}, \bibinfo{person}{Sandeep Soni},
  \bibinfo{person}{Jaime Teevan}, {and} \bibinfo{person}{Andr{\'e}s
  Monroy-Hern{\'a}ndez}.} \bibinfo{year}{2017}\natexlab{}.
\newblock \showarticletitle{Calendar. help: Designing a workflow-based
  scheduling agent with humans in the loop}. In
  \bibinfo{booktitle}{\emph{Proceedings of the 2017 CHI Conference on Human
  Factors in Computing Systems}}. ACM, \bibinfo{pages}{2382--2393}.
\newblock


\bibitem[\protect\citeauthoryear{Dang, Jin, et~al\mbox{.}}{Dang
  et~al\mbox{.}}{2018}]%
        {dang2018predict}
\bibfield{author}{\bibinfo{person}{Tommy Dang}, \bibinfo{person}{Fang Jin},
  {et~al\mbox{.}}} \bibinfo{year}{2018}\natexlab{}.
\newblock \showarticletitle{Predict saturated thickness using tensorboard
  visualization}. In \bibinfo{booktitle}{\emph{Proceedings of the Workshop on
  Visualisation in Environmental Sciences}}. Eurographics Association,
  \bibinfo{pages}{35--39}.
\newblock


\bibitem[\protect\citeauthoryear{DataRobot}{DataRobot}{[n.d.]}]%
        {datarobot}
\bibfield{author}{\bibinfo{person}{DataRobot}.}
  \bibinfo{year}{[n.d.]}\natexlab{}.
\newblock \bibinfo{title}{Automated Machine Learning for Predictive Modeling}.
\newblock
\newblock
\urldef\tempurl%
\url{https://www.datarobot.com/}
\showURL{%
Retrieved 3-April-2019 from \tempurl}


\bibitem[\protect\citeauthoryear{Donoho}{Donoho}{2017}]%
        {donoho201750}
\bibfield{author}{\bibinfo{person}{David Donoho}.}
  \bibinfo{year}{2017}\natexlab{}.
\newblock \showarticletitle{50 years of data science}.
\newblock \bibinfo{journal}{\emph{Journal of Computational and Graphical
  Statistics}} \bibinfo{volume}{26}, \bibinfo{number}{4}
  (\bibinfo{year}{2017}), \bibinfo{pages}{745--766}.
\newblock


\bibitem[\protect\citeauthoryear{Dourish and Bellotti}{Dourish and
  Bellotti}{1992}]%
        {dourish1992awareness}
\bibfield{author}{\bibinfo{person}{Paul Dourish} {and}
  \bibinfo{person}{Victoria Bellotti}.} \bibinfo{year}{1992}\natexlab{}.
\newblock \showarticletitle{Awareness and coordination in shared workspaces.}.
  In \bibinfo{booktitle}{\emph{CSCW}}, Vol.~\bibinfo{volume}{92}.
  \bibinfo{pages}{107--114}.
\newblock


\bibitem[\protect\citeauthoryear{Drozdal}{Drozdal}{2020}]%
        {JaimieDrozdal2020}
\bibfield{author}{\bibinfo{person}{Jaimie et~al. Drozdal}.}
  \bibinfo{year}{2020}\natexlab{}.
\newblock \showarticletitle{Exploring Information Needs for Establishing Trust
  in Automated Data Science Systems}. In \bibinfo{booktitle}{\emph{IUI'20}}.
  ACM, \bibinfo{pages}{in press}.
\newblock


\bibitem[\protect\citeauthoryear{EpistasisLab}{EpistasisLab}{[n.d.]}]%
        {web:tpot}
\bibfield{author}{\bibinfo{person}{EpistasisLab}.}
  \bibinfo{year}{[n.d.]}\natexlab{}.
\newblock \bibinfo{booktitle}{\emph{tpot}}.
\newblock
\urldef\tempurl%
\url{https://github.com/EpistasisLab/tpot}
\showURL{%
Retrieved 3-April-2019 from \tempurl}


\bibitem[\protect\citeauthoryear{Erickson and Kellogg}{Erickson and
  Kellogg}{2000}]%
        {erickson2000social}
\bibfield{author}{\bibinfo{person}{Thomas Erickson} {and}
  \bibinfo{person}{Wendy~A Kellogg}.} \bibinfo{year}{2000}\natexlab{}.
\newblock \showarticletitle{Social translucence: an approach to designing
  systems that support social processes}.
\newblock \bibinfo{journal}{\emph{ACM transactions on computer-human
  interaction (TOCHI)}} \bibinfo{volume}{7}, \bibinfo{number}{1}
  (\bibinfo{year}{2000}), \bibinfo{pages}{59--83}.
\newblock


\bibitem[\protect\citeauthoryear{Feurer, Klein, Eggensperger, Springenberg,
  Blum, and Hutter}{Feurer et~al\mbox{.}}{2015}]%
        {feurer2015efficient}
\bibfield{author}{\bibinfo{person}{Matthias Feurer}, \bibinfo{person}{Aaron
  Klein}, \bibinfo{person}{Katharina Eggensperger}, \bibinfo{person}{Jost
  Springenberg}, \bibinfo{person}{Manuel Blum}, {and} \bibinfo{person}{Frank
  Hutter}.} \bibinfo{year}{2015}\natexlab{}.
\newblock \showarticletitle{Efficient and robust automated machine learning}.
  In \bibinfo{booktitle}{\emph{Advances in Neural Information Processing
  Systems}}. \bibinfo{pages}{2962--2970}.
\newblock


\bibitem[\protect\citeauthoryear{Gergle and Tan}{Gergle and Tan}{2014}]%
        {gergle2014experimental}
\bibfield{author}{\bibinfo{person}{Darren Gergle} {and}
  \bibinfo{person}{Desney~S Tan}.} \bibinfo{year}{2014}\natexlab{}.
\newblock \showarticletitle{Experimental research in HCI}.
\newblock In \bibinfo{booktitle}{\emph{Ways of Knowing in HCI}}.
  \bibinfo{publisher}{Springer}, \bibinfo{pages}{191--227}.
\newblock


\bibitem[\protect\citeauthoryear{Gil, Honaker, Gupta, Ma, D'Orazio, Garijo,
  Gadewar, Yang, and Jahanshad}{Gil et~al\mbox{.}}{2019}]%
        {gil2019towards}
\bibfield{author}{\bibinfo{person}{Yolanda Gil}, \bibinfo{person}{James
  Honaker}, \bibinfo{person}{Shikhar Gupta}, \bibinfo{person}{Yibo Ma},
  \bibinfo{person}{Vito D'Orazio}, \bibinfo{person}{Daniel Garijo},
  \bibinfo{person}{Shruti Gadewar}, \bibinfo{person}{Qifan Yang}, {and}
  \bibinfo{person}{Neda Jahanshad}.} \bibinfo{year}{2019}\natexlab{}.
\newblock \showarticletitle{Towards human-guided machine learning}. In
  \bibinfo{booktitle}{\emph{Proceedings of the 24th International Conference on
  Intelligent User Interfaces}}. ACM, \bibinfo{pages}{614--624}.
\newblock


\bibitem[\protect\citeauthoryear{Google}{Google}{[n.d.]a}]%
        {web:googleautoml}
\bibfield{author}{\bibinfo{person}{Google}.}
  \bibinfo{year}{[n.d.]}\natexlab{a}.
\newblock \bibinfo{booktitle}{\emph{Cloud AutoML}}.
\newblock
\urldef\tempurl%
\url{https://cloud.google.com/automl/}
\showURL{%
Retrieved 3-April-2019 from \tempurl}


\bibitem[\protect\citeauthoryear{Google}{Google}{[n.d.]b}]%
        {web:colab}
\bibfield{author}{\bibinfo{person}{Google}.}
  \bibinfo{year}{[n.d.]}\natexlab{b}.
\newblock \bibinfo{booktitle}{\emph{Colaboratory}}.
\newblock
\urldef\tempurl%
\url{https://colab.research.google.com}
\showURL{%
Retrieved 3-April-2019 from \tempurl}


\bibitem[\protect\citeauthoryear{Grudin}{Grudin}{2009}]%
        {grudin2009ai}
\bibfield{author}{\bibinfo{person}{Jonathan Grudin}.}
  \bibinfo{year}{2009}\natexlab{}.
\newblock \showarticletitle{AI and HCI: Two fields divided by a common focus}.
\newblock \bibinfo{journal}{\emph{Ai Magazine}} \bibinfo{volume}{30},
  \bibinfo{number}{4} (\bibinfo{year}{2009}), \bibinfo{pages}{48--48}.
\newblock


\bibitem[\protect\citeauthoryear{Guo, Kandel, Hellerstein, and Heer}{Guo
  et~al\mbox{.}}{2011}]%
        {guo2011proactive}
\bibfield{author}{\bibinfo{person}{Philip~J Guo}, \bibinfo{person}{Sean
  Kandel}, \bibinfo{person}{Joseph~M Hellerstein}, {and}
  \bibinfo{person}{Jeffrey Heer}.} \bibinfo{year}{2011}\natexlab{}.
\newblock \showarticletitle{Proactive wrangling: mixed-initiative end-user
  programming of data transformation scripts}. In
  \bibinfo{booktitle}{\emph{Proceedings of the 24th annual ACM symposium on
  User interface software and technology}}. ACM, \bibinfo{pages}{65--74}.
\newblock


\bibitem[\protect\citeauthoryear{H2O}{H2O}{[n.d.]}]%
        {web:h2o}
\bibfield{author}{\bibinfo{person}{H2O}.} \bibinfo{year}{[n.d.]}\natexlab{}.
\newblock \bibinfo{booktitle}{\emph{H2O}}.
\newblock
\urldef\tempurl%
\url{https://h2o.ai}
\showURL{%
Retrieved 3-April-2019 from \tempurl}


\bibitem[\protect\citeauthoryear{He, Lin, Liu, Wang, Li, and Han}{He
  et~al\mbox{.}}{2018}]%
        {he2018amc}
\bibfield{author}{\bibinfo{person}{Yihui He}, \bibinfo{person}{Ji Lin},
  \bibinfo{person}{Zhijian Liu}, \bibinfo{person}{Hanrui Wang},
  \bibinfo{person}{Li-Jia Li}, {and} \bibinfo{person}{Song Han}.}
  \bibinfo{year}{2018}\natexlab{}.
\newblock \showarticletitle{Amc: Automl for model compression and acceleration
  on mobile devices}. In \bibinfo{booktitle}{\emph{Proceedings of the European
  Conference on Computer Vision (ECCV)}}. \bibinfo{pages}{784--800}.
\newblock


\bibitem[\protect\citeauthoryear{Heer and Shneiderman}{Heer and
  Shneiderman}{2012}]%
        {heer2012interactive}
\bibfield{author}{\bibinfo{person}{Jeffrey Heer} {and} \bibinfo{person}{Ben
  Shneiderman}.} \bibinfo{year}{2012}\natexlab{}.
\newblock \showarticletitle{Interactive dynamics for visual analysis}.
\newblock \bibinfo{journal}{\emph{Queue}} \bibinfo{volume}{10},
  \bibinfo{number}{2} (\bibinfo{year}{2012}), \bibinfo{pages}{30}.
\newblock


\bibitem[\protect\citeauthoryear{Heer, Vi{\'e}gas, and Wattenberg}{Heer
  et~al\mbox{.}}{2007}]%
        {heer2007voyagers}
\bibfield{author}{\bibinfo{person}{Jeffrey Heer}, \bibinfo{person}{Fernanda~B
  Vi{\'e}gas}, {and} \bibinfo{person}{Martin Wattenberg}.}
  \bibinfo{year}{2007}\natexlab{}.
\newblock \showarticletitle{Voyagers and voyeurs: supporting asynchronous
  collaborative information visualization}. In
  \bibinfo{booktitle}{\emph{Proceedings of the SIGCHI conference on Human
  factors in computing systems}}. ACM, \bibinfo{pages}{1029--1038}.
\newblock


\bibitem[\protect\citeauthoryear{Hoffman, Mueller, Klein, and Litman}{Hoffman
  et~al\mbox{.}}{2018}]%
        {hoffman2018metrics}
\bibfield{author}{\bibinfo{person}{Robert~R Hoffman}, \bibinfo{person}{Shane~T
  Mueller}, \bibinfo{person}{Gary Klein}, {and} \bibinfo{person}{Jordan
  Litman}.} \bibinfo{year}{2018}\natexlab{}.
\newblock \showarticletitle{Metrics for explainable AI: Challenges and
  prospects}.
\newblock \bibinfo{journal}{\emph{arXiv preprint arXiv:1812.04608}}
  (\bibinfo{year}{2018}).
\newblock


\bibitem[\protect\citeauthoryear{Horvitz}{Horvitz}{1999}]%
        {horvitz1999principles}
\bibfield{author}{\bibinfo{person}{Eric Horvitz}.}
  \bibinfo{year}{1999}\natexlab{}.
\newblock \showarticletitle{Principles of mixed-initiative user interfaces}. In
  \bibinfo{booktitle}{\emph{Proceedings of the SIGCHI conference on Human
  Factors in Computing Systems}}. \bibinfo{pages}{159--166}.
\newblock


\bibitem[\protect\citeauthoryear{Hou and Wang}{Hou and Wang}{2017}]%
        {hou2017hacking}
\bibfield{author}{\bibinfo{person}{Youyang Hou} {and} \bibinfo{person}{Dakuo
  Wang}.} \bibinfo{year}{2017}\natexlab{}.
\newblock \showarticletitle{Hacking with NPOs: collaborative analytics and
  broker roles in civic data hackathons}.
\newblock \bibinfo{journal}{\emph{Proceedings of the ACM on Human-Computer
  Interaction}} \bibinfo{volume}{1}, \bibinfo{number}{CSCW}
  (\bibinfo{year}{2017}), \bibinfo{pages}{53}.
\newblock


\bibitem[\protect\citeauthoryear{Jupyter}{Jupyter}{[n.d.]a}]%
        {web:jupyter}
\bibfield{author}{\bibinfo{person}{Project Jupyter}.}
  \bibinfo{year}{[n.d.]}\natexlab{a}.
\newblock \bibinfo{booktitle}{\emph{Jupyter Notebook}}.
\newblock
\urldef\tempurl%
\url{https://jupyter.org}
\showURL{%
Retrieved 3-April-2019 from \tempurl}


\bibitem[\protect\citeauthoryear{Jupyter}{Jupyter}{[n.d.]b}]%
        {web:jupyterlab}
\bibfield{author}{\bibinfo{person}{Project Jupyter}.}
  \bibinfo{year}{[n.d.]}\natexlab{b}.
\newblock \bibinfo{booktitle}{\emph{JupyterLab}}.
\newblock
\urldef\tempurl%
\url{https://www.github.com/jupyterlab/jupyterlab}
\showURL{%
\tempurl}


\bibitem[\protect\citeauthoryear{Kaggle}{Kaggle}{2017}]%
        {Kaggle2017}
\bibfield{author}{\bibinfo{person}{Kaggle}.} \bibinfo{year}{2017}\natexlab{}.
\newblock \bibinfo{booktitle}{\emph{The State of Data Science \& Machine
  Learning}}.
\newblock
\urldef\tempurl%
\url{https://www.kaggle.com/kaggle/kaggle-survey-2017}
\showURL{%
\tempurl}


\bibitem[\protect\citeauthoryear{Kaggle}{Kaggle}{2018}]%
        {kaggle2018survey}
\bibfield{author}{\bibinfo{person}{Kaggle}.} \bibinfo{year}{2018}\natexlab{}.
\newblock \bibinfo{booktitle}{\emph{Kaggle Data Science Survey 2018}}.
\newblock
\urldef\tempurl%
\url{https://www.kaggle.com/sudhirnl7/data-science-survey-2018/}
\showURL{%
Retrieved 17-September-2019 from \tempurl}


\bibitem[\protect\citeauthoryear{Kandel, Paepcke, Hellerstein, and Heer}{Kandel
  et~al\mbox{.}}{2011}]%
        {kandel2011wrangler}
\bibfield{author}{\bibinfo{person}{Sean Kandel}, \bibinfo{person}{Andreas
  Paepcke}, \bibinfo{person}{Joseph Hellerstein}, {and}
  \bibinfo{person}{Jeffrey Heer}.} \bibinfo{year}{2011}\natexlab{}.
\newblock \showarticletitle{Wrangler: Interactive visual specification of data
  transformation scripts}. In \bibinfo{booktitle}{\emph{Proceedings of the
  SIGCHI Conference on Human Factors in Computing Systems}}. ACM,
  \bibinfo{pages}{3363--3372}.
\newblock


\bibitem[\protect\citeauthoryear{Kanter and Veeramachaneni}{Kanter and
  Veeramachaneni}{2015}]%
        {kanter2015deep}
\bibfield{author}{\bibinfo{person}{James~Max Kanter} {and}
  \bibinfo{person}{Kalyan Veeramachaneni}.} \bibinfo{year}{2015}\natexlab{}.
\newblock \showarticletitle{Deep feature synthesis: Towards automating data
  science endeavors}. In \bibinfo{booktitle}{\emph{2015 IEEE International
  Conference on Data Science and Advanced Analytics (DSAA)}}. IEEE,
  \bibinfo{pages}{1--10}.
\newblock


\bibitem[\protect\citeauthoryear{Kay, Patel, and Kientz}{Kay
  et~al\mbox{.}}{2015}]%
        {kay2015good}
\bibfield{author}{\bibinfo{person}{Matthew Kay}, \bibinfo{person}{Shwetak~N
  Patel}, {and} \bibinfo{person}{Julie~A Kientz}.}
  \bibinfo{year}{2015}\natexlab{}.
\newblock \showarticletitle{How good is 85\%? A survey tool to connect
  classifier evaluation to acceptability of accuracy}. In
  \bibinfo{booktitle}{\emph{Proceedings of the 33rd Annual ACM Conference on
  Human Factors in Computing Systems}}. \bibinfo{pages}{347--356}.
\newblock


\bibitem[\protect\citeauthoryear{Kery, Radensky, Arya, John, and Myers}{Kery
  et~al\mbox{.}}{2018}]%
        {kery2018story}
\bibfield{author}{\bibinfo{person}{Mary~Beth Kery}, \bibinfo{person}{Marissa
  Radensky}, \bibinfo{person}{Mahima Arya}, \bibinfo{person}{Bonnie~E John},
  {and} \bibinfo{person}{Brad~A Myers}.} \bibinfo{year}{2018}\natexlab{}.
\newblock \showarticletitle{The story in the notebook: Exploratory data science
  using a literate programming tool}. In \bibinfo{booktitle}{\emph{Proceedings
  of the 2018 CHI Conference on Human Factors in Computing Systems}}. ACM,
  \bibinfo{pages}{174}.
\newblock


\bibitem[\protect\citeauthoryear{Khurana, Turaga, Samulowitz, and
  Parthasrathy}{Khurana et~al\mbox{.}}{2016}]%
        {khurana2016cognito}
\bibfield{author}{\bibinfo{person}{Udayan Khurana}, \bibinfo{person}{Deepak
  Turaga}, \bibinfo{person}{Horst Samulowitz}, {and}
  \bibinfo{person}{Srinivasan Parthasrathy}.} \bibinfo{year}{2016}\natexlab{}.
\newblock \showarticletitle{Cognito: Automated feature engineering for
  supervised learning}. In \bibinfo{booktitle}{\emph{2016 IEEE 16th
  International Conference on Data Mining Workshops (ICDMW)}}. IEEE,
  \bibinfo{pages}{1304--1307}.
\newblock


\bibitem[\protect\citeauthoryear{Krensky, Harner, Brethenoux, Hare, Sicular,
  and Vashisth}{Krensky et~al\mbox{.}}{2020}]%
        {gartner2020magic}
\bibfield{author}{\bibinfo{person}{Peter Krensky}, \bibinfo{person}{Pieter~den
  Harner}, \bibinfo{person}{Erick Brethenoux}, \bibinfo{person}{Jim Hare},
  \bibinfo{person}{Svetlana Sicular}, {and} \bibinfo{person}{Shubhangi
  Vashisth}.} \bibinfo{year}{2020}\natexlab{}.
\newblock \showarticletitle{Magic Quadrant for data science and
  machine-learning platforms}.
\newblock \bibinfo{journal}{\emph{Gartner, Inc}} (\bibinfo{year}{2020}).
\newblock


\bibitem[\protect\citeauthoryear{Lam, Thiebaut, Sinn, Chen, Mai, and Alkan}{Lam
  et~al\mbox{.}}{2017}]%
        {lam2017one}
\bibfield{author}{\bibinfo{person}{Hoang~Thanh Lam},
  \bibinfo{person}{Johann-Michael Thiebaut}, \bibinfo{person}{Mathieu Sinn},
  \bibinfo{person}{Bei Chen}, \bibinfo{person}{Tiep Mai}, {and}
  \bibinfo{person}{Oznur Alkan}.} \bibinfo{year}{2017}\natexlab{}.
\newblock \showarticletitle{One button machine for automating feature
  engineering in relational databases}.
\newblock \bibinfo{journal}{\emph{arXiv preprint arXiv:1706.00327}}
  (\bibinfo{year}{2017}).
\newblock


\bibitem[\protect\citeauthoryear{Lee, Macke, Xin, Lee, Huang, and
  Parameswaran}{Lee et~al\mbox{.}}{2019}]%
        {lee2019human}
\bibfield{author}{\bibinfo{person}{Doris Jung-Lin Lee},
  \bibinfo{person}{Stephen Macke}, \bibinfo{person}{Doris Xin},
  \bibinfo{person}{Angela Lee}, \bibinfo{person}{Silu Huang}, {and}
  \bibinfo{person}{Aditya Parameswaran}.} \bibinfo{year}{2019}\natexlab{}.
\newblock \showarticletitle{A Human-in-the-loop Perspective on AutoML:
  Milestones and the Road Ahead}.
\newblock \bibinfo{journal}{\emph{Data Engineering}} (\bibinfo{year}{2019}),
  \bibinfo{pages}{58}.
\newblock


\bibitem[\protect\citeauthoryear{Li}{Li}{2018}]%
        {li2018goodai}
\bibfield{author}{\bibinfo{person}{Fei-Fei Li}.}
  \bibinfo{year}{2018}\natexlab{}.
\newblock \showarticletitle{How to Make A.I. That's Good for People}.
\newblock \bibinfo{journal}{\emph{The New York Times}} (\bibinfo{date}{7 March}
  \bibinfo{year}{2018}).
\newblock
\urldef\tempurl%
\url{https://www.nytimes.com/2018/03/07/opinion/artificial-intelligence-human.html}
\showURL{%
Retrieved 3-April-2019 from \tempurl}


\bibitem[\protect\citeauthoryear{Liu, Ram, Vijaykeerthy, Bouneffouf, Bramble,
  Samulowitz, Wang, Conn, and Gray}{Liu et~al\mbox{.}}{2020}]%
        {liu2020admm}
\bibfield{author}{\bibinfo{person}{Sijia Liu}, \bibinfo{person}{Parikshit Ram},
  \bibinfo{person}{Deepak Vijaykeerthy}, \bibinfo{person}{Djallel Bouneffouf},
  \bibinfo{person}{Gregory Bramble}, \bibinfo{person}{Horst Samulowitz},
  \bibinfo{person}{Dakuo Wang}, \bibinfo{person}{Andrew Conn}, {and}
  \bibinfo{person}{Alexander~G Gray}.} \bibinfo{year}{2020}\natexlab{}.
\newblock \showarticletitle{An ADMM Based Framework for AutoML Pipeline
  Configuration.}. In \bibinfo{booktitle}{\emph{AAAI}}.
  \bibinfo{pages}{4892--4899}.
\newblock


\bibitem[\protect\citeauthoryear{Mao, Wang, Muller, Varshney, Baldini, Dugan,
  and Mojsilovic}{Mao et~al\mbox{.}}{2020}]%
        {mao2019}
\bibfield{author}{\bibinfo{person}{Yaoli Mao}, \bibinfo{person}{Dakuo Wang},
  \bibinfo{person}{Michael Muller}, \bibinfo{person}{Kush Varshney},
  \bibinfo{person}{Ioana Baldini}, \bibinfo{person}{Casey Dugan}, {and}
  \bibinfo{person}{Aleksandra Mojsilovic}.} \bibinfo{year}{2020}\natexlab{}.
\newblock \showarticletitle{How Data Scientists Work Together With Domain
  Experts in Scientific Collaborations}. In
  \bibinfo{booktitle}{\emph{Proceedings of the 2020 ACM conference on GROUP}}.
  ACM.
\newblock


\bibitem[\protect\citeauthoryear{Markoff}{Markoff}{2011}]%
        {markoff2011computer}
\bibfield{author}{\bibinfo{person}{John Markoff}.}
  \bibinfo{year}{2011}\natexlab{}.
\newblock \showarticletitle{Computer wins on jeopardy!: trivial, it is not}.
\newblock \bibinfo{journal}{\emph{New York Times}}  \bibinfo{volume}{16}
  (\bibinfo{year}{2011}).
\newblock


\bibitem[\protect\citeauthoryear{McDonald}{McDonald}{2009}]%
        {mcdonald2009handbook}
\bibfield{author}{\bibinfo{person}{John~H McDonald}.}
  \bibinfo{year}{2009}\natexlab{}.
\newblock \bibinfo{booktitle}{\emph{Handbook of biological statistics}}.
  Vol.~\bibinfo{volume}{2}.
\newblock \bibinfo{publisher}{sparky house publishing Baltimore, MD}.
\newblock


\bibitem[\protect\citeauthoryear{Miles and Shevlin}{Miles and Shevlin}{2001}]%
        {miles2001applying}
\bibfield{author}{\bibinfo{person}{Jeremy Miles} {and} \bibinfo{person}{Mark
  Shevlin}.} \bibinfo{year}{2001}\natexlab{}.
\newblock \bibinfo{booktitle}{\emph{Applying regression and correlation: A
  guide for students and researchers}}.
\newblock \bibinfo{publisher}{Sage}.
\newblock


\bibitem[\protect\citeauthoryear{Muller, Lange, Wang, Piorkowski, Tsay, Liao,
  Dugan, and Erickson}{Muller et~al\mbox{.}}{2019}]%
        {muller2019datascience}
\bibfield{author}{\bibinfo{person}{Michael Muller}, \bibinfo{person}{Ingrid
  Lange}, \bibinfo{person}{Dakuo Wang}, \bibinfo{person}{David Piorkowski},
  \bibinfo{person}{Jason Tsay}, \bibinfo{person}{Q.~Vera Liao},
  \bibinfo{person}{Casey Dugan}, {and} \bibinfo{person}{Thomas Erickson}.}
  \bibinfo{year}{2019}\natexlab{}.
\newblock \showarticletitle{How Data Science Workers Work with Data: Discovery,
  Capture, Curation, Design, Creation}. In
  \bibinfo{booktitle}{\emph{Proceedings of the 2019 CHI Conference on Human
  Factors in Computing Systems}} (Glasgow, UK) \emph{(\bibinfo{series}{CHI
  '19})}. \bibinfo{publisher}{ACM}, \bibinfo{address}{New York, NY, USA},
  \bibinfo{pages}{Forthcoming}.
\newblock


\bibitem[\protect\citeauthoryear{Nass, Steuer, and Tauber}{Nass
  et~al\mbox{.}}{1994}]%
        {nass1994computers}
\bibfield{author}{\bibinfo{person}{Clifford Nass}, \bibinfo{person}{Jonathan
  Steuer}, {and} \bibinfo{person}{Ellen~R Tauber}.}
  \bibinfo{year}{1994}\natexlab{}.
\newblock \showarticletitle{Computers are social actors}. In
  \bibinfo{booktitle}{\emph{Proceedings of the SIGCHI conference on Human
  factors in computing systems}}. ACM, \bibinfo{pages}{72--78}.
\newblock


\bibitem[\protect\citeauthoryear{Olson and Moore}{Olson and Moore}{2016}]%
        {olson2016tpot}
\bibfield{author}{\bibinfo{person}{Randal~S Olson} {and}
  \bibinfo{person}{Jason~H Moore}.} \bibinfo{year}{2016}\natexlab{}.
\newblock \showarticletitle{TPOT: A tree-based pipeline optimization tool for
  automating machine learning}. In \bibinfo{booktitle}{\emph{Workshop on
  Automatic Machine Learning}}. \bibinfo{pages}{66--74}.
\newblock


\bibitem[\protect\citeauthoryear{Pedregosa, Varoquaux, Gramfort, Michel,
  Thirion, Grisel, Blondel, Prettenhofer, Weiss, Dubourg,
  et~al\mbox{.}}{Pedregosa et~al\mbox{.}}{2011}]%
        {pedregosa2011scikit}
\bibfield{author}{\bibinfo{person}{Fabian Pedregosa}, \bibinfo{person}{Ga{\"e}l
  Varoquaux}, \bibinfo{person}{Alexandre Gramfort}, \bibinfo{person}{Vincent
  Michel}, \bibinfo{person}{Bertrand Thirion}, \bibinfo{person}{Olivier
  Grisel}, \bibinfo{person}{Mathieu Blondel}, \bibinfo{person}{Peter
  Prettenhofer}, \bibinfo{person}{Ron Weiss}, \bibinfo{person}{Vincent
  Dubourg}, {et~al\mbox{.}}} \bibinfo{year}{2011}\natexlab{}.
\newblock \showarticletitle{Scikit-learn: Machine learning in Python}.
\newblock \bibinfo{journal}{\emph{Journal of machine learning research}}
  \bibinfo{volume}{12}, \bibinfo{number}{Oct} (\bibinfo{year}{2011}),
  \bibinfo{pages}{2825--2830}.
\newblock


\bibitem[\protect\citeauthoryear{Piorkowski, Park, Wang, Wang, Muller, and
  Portnoy}{Piorkowski et~al\mbox{.}}{2021}]%
        {dscommunicate}
\bibfield{author}{\bibinfo{person}{David Piorkowski}, \bibinfo{person}{Soya
  Park}, \bibinfo{person}{April~Yi Wang}, \bibinfo{person}{Dakuo Wang},
  \bibinfo{person}{Michael Muller}, {and} \bibinfo{person}{Felix Portnoy}.}
  \bibinfo{year}{2021}\natexlab{}.
\newblock \showarticletitle{How AI Developers Overcome Communication Challenges
  in a Multidisciplinary Team: A Case Study}. In
  \bibinfo{booktitle}{\emph{Proceedings of the CSCW 2021}}.
\newblock


\bibitem[\protect\citeauthoryear{Rattenbury, Hellerstein, Heer, Kandel, and
  Carreras}{Rattenbury et~al\mbox{.}}{2017}]%
        {rattenbury2017principles}
\bibfield{author}{\bibinfo{person}{Tye Rattenbury}, \bibinfo{person}{Joseph~M
  Hellerstein}, \bibinfo{person}{Jeffrey Heer}, \bibinfo{person}{Sean Kandel},
  {and} \bibinfo{person}{Connor Carreras}.} \bibinfo{year}{2017}\natexlab{}.
\newblock \bibinfo{booktitle}{\emph{Principles of data wrangling: Practical
  techniques for data preparation}}.
\newblock \bibinfo{publisher}{" O'Reilly Media, Inc."}.
\newblock


\bibitem[\protect\citeauthoryear{Roy, Zhang, and Vogel}{Roy
  et~al\mbox{.}}{2019}]%
        {roy2019automation}
\bibfield{author}{\bibinfo{person}{Quentin Roy}, \bibinfo{person}{Futian
  Zhang}, {and} \bibinfo{person}{Daniel Vogel}.}
  \bibinfo{year}{2019}\natexlab{}.
\newblock \showarticletitle{Automation accuracy is good, but high
  controllability may be better}. In \bibinfo{booktitle}{\emph{Proceedings of
  the 2019 CHI Conference on Human Factors in Computing Systems}}.
  \bibinfo{pages}{1--8}.
\newblock


\bibitem[\protect\citeauthoryear{Ruiz}{Ruiz}{2017}]%
        {ruiz201780}
\bibfield{author}{\bibinfo{person}{Armand Ruiz}.}
  \bibinfo{year}{2017}\natexlab{}.
\newblock \showarticletitle{The 80/20 data science dilemma}.
\newblock \bibinfo{journal}{\emph{InfoWorld}} (\bibinfo{year}{2017}).
\newblock


\bibitem[\protect\citeauthoryear{Schwab}{Schwab}{2018}]%
        {schwab_2018}
\bibfield{author}{\bibinfo{person}{Katharine Schwab}.}
  \bibinfo{year}{2018}\natexlab{}.
\newblock \bibinfo{title}{Google's Rules For Designers Working With AI}.
\newblock
\newblock
\urldef\tempurl%
\url{https://www.fastcompany.com/90132700/googles-rules-for-designing-ai-that-isnt-evil}
\showURL{%
\tempurl}


\bibitem[\protect\citeauthoryear{Shamekhi, Liao, Wang, Bellamy, and
  Erickson}{Shamekhi et~al\mbox{.}}{2018}]%
        {shamekhi2018face}
\bibfield{author}{\bibinfo{person}{Ameneh Shamekhi}, \bibinfo{person}{Q~Vera
  Liao}, \bibinfo{person}{Dakuo Wang}, \bibinfo{person}{Rachel~KE Bellamy},
  {and} \bibinfo{person}{Thomas Erickson}.} \bibinfo{year}{2018}\natexlab{}.
\newblock \showarticletitle{Face Value? Exploring the effects of embodiment for
  a group facilitation agent}. In \bibinfo{booktitle}{\emph{Proceedings of the
  2018 CHI Conference on Human Factors in Computing Systems}}. ACM,
  \bibinfo{pages}{391}.
\newblock


\bibitem[\protect\citeauthoryear{Shneiderman and Maes}{Shneiderman and
  Maes}{1997}]%
        {shneiderman1997direct}
\bibfield{author}{\bibinfo{person}{Ben Shneiderman} {and}
  \bibinfo{person}{Pattie Maes}.} \bibinfo{year}{1997}\natexlab{}.
\newblock \showarticletitle{Direct manipulation vs. interface agents}.
\newblock \bibinfo{journal}{\emph{interactions}} \bibinfo{volume}{4},
  \bibinfo{number}{6} (\bibinfo{year}{1997}), \bibinfo{pages}{42--61}.
\newblock


\bibitem[\protect\citeauthoryear{Simon}{Simon}{2018}]%
        {simon_2018}
\bibfield{author}{\bibinfo{person}{Matt Simon}.}
  \bibinfo{year}{2018}\natexlab{}.
\newblock \bibinfo{title}{Forget the Robot Singularity Apocalypse. Let's Talk
  About the Multiplicity}.
\newblock
\newblock
\urldef\tempurl%
\url{https://www.wired.com/story/forget-the-robot-singularity-apocalypse-lets-talk-about-the-multiplicity/}
\showURL{%
\tempurl}


\bibitem[\protect\citeauthoryear{Sukis}{Sukis}{2019}]%
        {sukis_2019}
\bibfield{author}{\bibinfo{person}{Jennifer Sukis}.}
  \bibinfo{year}{2019}\natexlab{}.
\newblock \bibinfo{title}{AI Design \& Practices Guidelines}.
\newblock
\newblock
\urldef\tempurl%
\url{https://medium.com/design-ibm/ai-design-guidelines-e06f7e92d864}
\showURL{%
\tempurl}


\bibitem[\protect\citeauthoryear{Sutton, Hobson, Geddes, and Caruana}{Sutton
  et~al\mbox{.}}{2018}]%
        {sutton2018data}
\bibfield{author}{\bibinfo{person}{Charles Sutton}, \bibinfo{person}{Timothy
  Hobson}, \bibinfo{person}{James Geddes}, {and} \bibinfo{person}{Rich
  Caruana}.} \bibinfo{year}{2018}\natexlab{}.
\newblock \showarticletitle{Data diff: Interpretable, executable summaries of
  changes in distributions for data wrangling}. In
  \bibinfo{booktitle}{\emph{Proceedings of the 24th ACM SIGKDD International
  Conference on Knowledge Discovery \& Data Mining}}. ACM,
  \bibinfo{pages}{2279--2288}.
\newblock


\bibitem[\protect\citeauthoryear{Tan, Wang, and Sabanovic}{Tan
  et~al\mbox{.}}{2018}]%
        {tan2018projecting}
\bibfield{author}{\bibinfo{person}{Haodan Tan}, \bibinfo{person}{Dakuo Wang},
  {and} \bibinfo{person}{Selma Sabanovic}.} \bibinfo{year}{2018}\natexlab{}.
\newblock \showarticletitle{Projecting Life Onto Robots: The Effects of
  Cultural Factors and Design Type on Multi-Level Evaluations of Robot
  Anthropomorphism}. In \bibinfo{booktitle}{\emph{2018 27th IEEE International
  Symposium on Robot and Human Interactive Communication (RO-MAN)}}. IEEE,
  \bibinfo{pages}{129--136}.
\newblock


\bibitem[\protect\citeauthoryear{Wang, Liao, Zhang, Khurana, Samulowitz, Park,
  Muller, and Amini}{Wang et~al\mbox{.}}{2021a}]%
        {automationsurvey}
\bibfield{author}{\bibinfo{person}{Dakuo Wang}, \bibinfo{person}{Q.~Vera Liao},
  \bibinfo{person}{Yunfeng Zhang}, \bibinfo{person}{Udayan Khurana},
  \bibinfo{person}{Horst Samulowitz}, \bibinfo{person}{Soya Park},
  \bibinfo{person}{Michael Muller}, {and} \bibinfo{person}{Lisa Amini}.}
  \bibinfo{year}{2021}\natexlab{a}.
\newblock \showarticletitle{How Much Automation Does a Data Scientist Want?}.
  In \bibinfo{booktitle}{\emph{pre-print}}.
\newblock


\bibitem[\protect\citeauthoryear{Wang, Wang, Zhang, Wang, Zhu, Gao, Fan, and
  Tian}{Wang et~al\mbox{.}}{2021b}]%
        {aidoctor}
\bibfield{author}{\bibinfo{person}{Dakuo Wang}, \bibinfo{person}{Liuping Wang},
  \bibinfo{person}{Zhan Zhang}, \bibinfo{person}{Ding Wang},
  \bibinfo{person}{Haiyi Zhu}, \bibinfo{person}{Yvonne Gao},
  \bibinfo{person}{Xiangmin Fan}, {and} \bibinfo{person}{Feng Tian}.}
  \bibinfo{year}{2021}\natexlab{b}.
\newblock \showarticletitle{Brilliant AI Doctor in Rural China: Tensions and
  Challenges in AI-Powered CDSS Deployment}. In
  \bibinfo{booktitle}{\emph{Proceedings of the CHI 2021}}.
\newblock


\bibitem[\protect\citeauthoryear{Wang, Weisz, Muller, Ram, Geyer, Dugan,
  Tausczik, Samulowitz, and Gray}{Wang et~al\mbox{.}}{2019b}]%
        {wang2019humanai}
\bibfield{author}{\bibinfo{person}{Dakuo Wang}, \bibinfo{person}{Justin~D.
  Weisz}, \bibinfo{person}{Michael Muller}, \bibinfo{person}{Parikshit Ram},
  \bibinfo{person}{Werner Geyer}, \bibinfo{person}{Casey Dugan},
  \bibinfo{person}{Yla Tausczik}, \bibinfo{person}{Horst Samulowitz}, {and}
  \bibinfo{person}{Alexander Gray}.} \bibinfo{year}{2019}\natexlab{b}.
\newblock \showarticletitle{Human-AI Collaboration in Data Science: Exploring
  Data Scientists' Perceptions of Automated AI}.
\newblock \bibinfo{journal}{\emph{To appear in Computer Supported Cooperative
  Work (CSCW)}} (\bibinfo{year}{2019}).
\newblock


\bibitem[\protect\citeauthoryear{Wang, Zhang, Zheng, Wang, Yuan, Dai, Zhang,
  and Yang}{Wang et~al\mbox{.}}{2016}]%
        {wang2016does}
\bibfield{author}{\bibinfo{person}{Fei-Yue Wang}, \bibinfo{person}{Jun~Jason
  Zhang}, \bibinfo{person}{Xinhu Zheng}, \bibinfo{person}{Xiao Wang},
  \bibinfo{person}{Yong Yuan}, \bibinfo{person}{Xiaoxiao Dai},
  \bibinfo{person}{Jie Zhang}, {and} \bibinfo{person}{Liuqing Yang}.}
  \bibinfo{year}{2016}\natexlab{}.
\newblock \showarticletitle{Where does AlphaGo go: From church-turing thesis to
  AlphaGo thesis and beyond}.
\newblock \bibinfo{journal}{\emph{IEEE/CAA Journal of Automatica Sinica}}
  \bibinfo{volume}{3}, \bibinfo{number}{2} (\bibinfo{year}{2016}),
  \bibinfo{pages}{113--120}.
\newblock


\bibitem[\protect\citeauthoryear{Wang, Ming, Jin, Shen, Liu, Smith,
  Veeramachaneni, and Qu}{Wang et~al\mbox{.}}{2019a}]%
        {wang2019atmseer}
\bibfield{author}{\bibinfo{person}{Qianwen Wang}, \bibinfo{person}{Yao Ming},
  \bibinfo{person}{Zhihua Jin}, \bibinfo{person}{Qiaomu Shen},
  \bibinfo{person}{Dongyu Liu}, \bibinfo{person}{Micah~J Smith},
  \bibinfo{person}{Kalyan Veeramachaneni}, {and} \bibinfo{person}{Huamin Qu}.}
  \bibinfo{year}{2019}\natexlab{a}.
\newblock \showarticletitle{Atmseer: Increasing transparency and
  controllability in automated machine learning}. In
  \bibinfo{booktitle}{\emph{Proceedings of the 2019 CHI Conference on Human
  Factors in Computing Systems}}. ACM, \bibinfo{pages}{681}.
\newblock


\bibitem[\protect\citeauthoryear{Weidele, Weisz, Oduor, Muller, Andres, Gray,
  and Wang}{Weidele et~al\mbox{.}}{2020}]%
        {weidele2020autoaiviz}
\bibfield{author}{\bibinfo{person}{Daniel Karl~I Weidele},
  \bibinfo{person}{Justin~D Weisz}, \bibinfo{person}{Erick Oduor},
  \bibinfo{person}{Michael Muller}, \bibinfo{person}{Josh Andres},
  \bibinfo{person}{Alexander Gray}, {and} \bibinfo{person}{Dakuo Wang}.}
  \bibinfo{year}{2020}\natexlab{}.
\newblock \showarticletitle{AutoAIViz: opening the blackbox of automated
  artificial intelligence with conditional parallel coordinates}. In
  \bibinfo{booktitle}{\emph{Proceedings of the 25th International Conference on
  Intelligent User Interfaces}}. \bibinfo{pages}{308--312}.
\newblock


\bibitem[\protect\citeauthoryear{Xu, Wang, Collins, Lee, and Warschauer}{Xu
  et~al\mbox{.}}{[n.d.]}]%
        {xu161same}
\bibfield{author}{\bibinfo{person}{Ying Xu}, \bibinfo{person}{Dakuo Wang},
  \bibinfo{person}{Penelope Collins}, \bibinfo{person}{Hyelim Lee}, {and}
  \bibinfo{person}{Mark Warschauer}.} \bibinfo{year}{[n.d.]}\natexlab{}.
\newblock \showarticletitle{Same benefits, different communication patterns:
  Comparing Children's reading with a conversational agent vs. a human
  partner}.
\newblock \bibinfo{journal}{\emph{Computers \& Education}}
  \bibinfo{volume}{161} (\bibinfo{year}{[n.\,d.]}), \bibinfo{pages}{104059}.
\newblock


\bibitem[\protect\citeauthoryear{Zhang, Muller, and Wang}{Zhang
  et~al\mbox{.}}{2020}]%
        {zhang2020data}
\bibfield{author}{\bibinfo{person}{Amy~X Zhang}, \bibinfo{person}{Michael
  Muller}, {and} \bibinfo{person}{Dakuo Wang}.}
  \bibinfo{year}{2020}\natexlab{}.
\newblock \showarticletitle{How do data science workers collaborate? roles,
  workflows, and tools}.
\newblock \bibinfo{journal}{\emph{Proceedings of the ACM on Human-Computer
  Interaction}} \bibinfo{volume}{4}, \bibinfo{number}{CSCW1}
  (\bibinfo{year}{2020}), \bibinfo{pages}{1--23}.
\newblock


\bibitem[\protect\citeauthoryear{Z{\"o}ller and Huber}{Z{\"o}ller and
  Huber}{2019}]%
        {zoller2019survey}
\bibfield{author}{\bibinfo{person}{Marc-Andr{\'e} Z{\"o}ller} {and}
  \bibinfo{person}{Marco~F Huber}.} \bibinfo{year}{2019}\natexlab{}.
\newblock \showarticletitle{Survey on Automated Machine Learning}.
\newblock \bibinfo{journal}{\emph{arXiv preprint arXiv:1904.12054}}
  (\bibinfo{year}{2019}).
\newblock


\end{thebibliography}

\end{document}